\documentclass[fleqn,usenatbib]{mnras}

\usepackage{newtxtext,newtxmath}

\usepackage[T1]{fontenc}

\DeclareRobustCommand{\VAN}[3]{#2}
\let\VANthebibliography\thebibliography
\def\thebibliography{\DeclareRobustCommand{\VAN}[3]{##3}\VANthebibliography}

\usepackage{graphicx}	
\usepackage{amsmath}	
\usepackage{multirow}
\usepackage{subfig}

\outer\def\gtae {$\buildrel {\lower3pt\hbox{$>$}} \over 
{\lower2pt\hbox{$\sim$}} $}
\outer\def\ltae {$\buildrel {\lower3pt\hbox{$<$}} \over 
{\lower2pt\hbox{$\sim$}} $}
\newcommand{\Msun}{$M_{\odot}$}

\newcommand{\Rsun}{$R_{\odot}$}

\newcommand{\kep}{\it Kepler}
\newcommand{\tess}{\it TESS}

\newcommand{\cz}{\ensuremath{C_Z}}
\newcommand{\bz}{\ensuremath{\langle B_z\rangle}}
\newcommand{\nnz}{\ensuremath{\langle N_z\rangle}}

\title[Magnetic Fields on M dwarf UFRs]{The Puzzling Story of Flare Inactive Ultra Fast Rotating M dwarfs. I. Exploring their Magnetic Fields}

\author[L. Doyle et al.]{
Lauren Doyle,$^{1,2}$\thanks{E-mail: lauren.doyle@warwick.ac.uk}
Stefano Bagnulo,$^{3}$
Gavin Ramsay$^{3}$
J. Gerry Doyle$^{3}$
and Pasi Hakala$^{4}$
\\
$^{1}$Centre for Exoplanets and Habitability, University of Warwick, Coventry, CV4 7AL, UK \\
$^{2}$Department of Physics, University of Warwick, Coventry, CV4 7AL, UK\\
$^{3}$Armagh Observatory and Planetarium, College Hill, Armagh, BT61 9DG, UK\\
$^{4}$Finnish Centre for Astronomy with ESO (FINCA), Quantum, University of Turku, FI-20014, Finland
}

\date{Accepted 2022 February 10. Received 2022 February 10; in original form 2021 October 8}

\pubyear{2022}

\begin{document}
\label{firstpage}
\pagerange{\pageref{firstpage}--\pageref{lastpage}}
\maketitle

\begin{abstract}
Stars which are rapidly rotating are expected to show high levels of activity according to the activity-rotation relation. However, previous {\tess} studies have found Ultra Fast Rotating (UFR) M dwarfs with periods less than one day displaying low levels of flaring activity. As a result, in this study, we utilise VLT/FORS2 spectropolarimetric data of ten M dwarf UFR stars between spectral types $\sim$M2 - M6 all with $P_{\rm rot} <$ 1, to detect the presence of a magnetic field. We divide our sample into rotation period bins of equal size, with one star having many more flares in the {\tess} lightcurve than the other. We also provide an analysis of the long-term variability within our sample using{\tess} lightcurves taken during Cycles 1 and 3 (up to three years apart). We identify 605 flares from our sample which have energies between 2.0$\times$10$^{31}$ and 5.4$\times$10$^{34}$~erg. Although we find no significance difference in the flare rate between the Cycles, two of our targets display changes in their lightcurve morphology, potentially caused by a difference in the spot distribution. Overall, we find five stars (50\%) in our sample have a detectable magnetic field with strengths $\sim$1--2~kG. Of these five, four were the more flare active stars within the period bins with one being the less flare active star. It would appear the magnetic field strength may not be the answer to the lack of flaring activity and supersaturation or magnetic field configuration may play a role. However, it is clear the relationship between rotation and activity is more complex than a steady decrease over time. 
\end{abstract}




\section{Introduction}

\begin{table*}
\caption{The stellar properties of the ten low mass stars in our survey: we show the TIC ID \citep{stassun2018tess}; catalogue name; the {\tess} sectors and duration each star was observed;  RA and DEC; MV Spectral type ({\tt SIMBAD}); the stellar effective temperature; stellar age;  $T_{mag}$ \citep{stassun2018tess}; their rotation period (this work); RUWE and Distance \citep{gaia16, gaia18, gaiaedr3brown} and the quiescent luminosity \citep{doyle2019}. Stellar ages are taken from the following sources using estimates from moving groups and associations: $^a$ \citet{zhan2019complex}, $^b$ \citet{janson2017binaries}, $^c$ \citet{loyd2021hazmat}, $^d$ \citet{booth2021age} and $^e$ \citet{gagne2014banyan}.}

   \begin{center}
   \label{stellar_properties}
\resizebox{1.0\textwidth}{!}{
	\begin{tabular}{llrrrrcrcccccccc}
    \hline 
	TIC ID    & Name                & TESS Sector              & Obs. Length & RA      & Dec       & SpT  & T$_{\rm{eff}}$ & Age   & $T_{mag}$ & $P_{\rm rot}$ & RUWE & Distance  & $log(L_{star})$  \\
	          &                     &                          & (days)             & (J2000) & (J2000)   &      & (K)          & (Myr) &           & (days)       &  & (pc)      & (erg/s)          \\
	\hline
    158596311 & UCAC4 204-001345 	&   2 \& 29	               & 44.08                 & 22.1269 & -49.3528  & 4.1  & 3159 & $\sim$45$^b$ &  12.3    & 0.154     & 1.45     & 43.4     &  31.14   \\
    425937691 &	UCAC3 53-724        &   1, 2 \& 28	           & 69.96                 & 5.3666	 & -63.8525  & 5.5  & 2877 & $\sim$45$^a$ &  13.2    & 0.10      & 2.85	   & 43.9     &  30.80   \\
    206544316 &	UPM J0113-5939   	&   1,2,28 \& 29	       & 92.45                 & 18.4189 & -59.6598  & 3.7  & 3237 & $\sim$150$^a$ &  11.6	   & 0.322	   & 1.26     & 43.1     &  31.39   \\ 
    156002545 &	2MASS J0033-5116 	&   2 \& 29	               & 45.30                 & 8.3524	 & -51.2790  & 3.4  & 3318 & $\sim$45$^b$ &  11.4    & 0.353     & 1.51    & 41.4     &  31.42   \\
    248354845 &	GSC 04683-02117  	&   3 \& 30	               & 40.67                 & 17.8474 & -5.4273   & 3.5  & 3300 &  &  11.1	   & 0.522     & 1.64     & 36.7     &  31.47   \\ 
    229142295 &	2MASS J0146-5339 	&   2, 3, 29 \& 30	       & 87.78                 & 26.6237 & -53.6596  & 4.5  & 3131 &  &  11.2	   & 0.447     & 3.32     & 17.5     &  30.79   \\ 
    220539110 &	GSC 08859-00633  	&   1-3 \& 28-30           & 133.42                & 43.4470 & -61.5878  & 3.0  & 3538 &  &  9.9	   & 0.773     & 7.52     & 41.5     &  31.98   \\ 
    166808151 &	EXO 0235.2-5216  	&   2,3,29 \& 30	       & 88.64                 & 39.2163 & -52.0510  & 2.0  & 3510 & $\sim$45$^c$ &  9.9     & 0.740     & 1.26     & 38.9     &  31.95   \\
    141807839 &	AL 442           	&   2-6,8,11-13 \& 27-38   & 465.24                & 92.8752 & -72.2271  & 4.5  & 3259 & $\sim$13$^d$ &  11.2    & 0.839     & --     & 56.9     &  31.81   \\
    201861769 &	2MASS J0232-5746 	&   2 \& 29	               & 46.51                 & 38.0806 & -57.7699  & 4.1  & 3134 & $\sim$45$^e$ &  12.8    & 0.862 	   & 1.29     & 45.9     &  30.97   \\
    \hline
    \end{tabular}}
    \end{center}
\end{table*}

{\kep} and {\tess} have provided high cadence photometric lightcurves of thousands of low mass stars, which has facilitated the study of rotation periods \citep[e.g.][]{mcquillan2013measuring, martins2020search, howard2020evryflare}, stellar flares \citep[e.g.][]{hawley2014kepler, davenport2016kepler, vida2019flaring, gunther2020stellar, feinstein2020flare} and magnetic activity \citep[e.g.][]{dmitrienko2017spots, vida2017frequent, seli2021activity, metcalfe2021magnetic}. Much of this interest has stemmed from the fact that low mass stars have been targeted in the search for exoplanets which are easier to detect around small stars \citep[e.g.][]{NutzmanCharbonneau2008}. In turn, understanding the magnetic activity of these stars is important in predicting the resulting effects of the host star on the atmosphere of any orbiting exoplanets \citep[e.g.][]{Badhan2019, vida2017frequent}. \citet{Skumanich1972} was amongst the first to show that as stars age they gradually spin down over time. This happens as the star loses angular momentum through stellar winds, reducing the stellar rotational velocity. As a result of this, magnetic activity (such as starspots and flares) is observed to decline over time for low mass main sequence stars. Therefore, stars which are rapidly rotating are expected to be young and show high levels of activity which is strongly related to their dynamo mechanism \citep[e.g.][]{hartmann1987rotation,maggio1987einstein}. This was recently demonstrated by \citet{davenport2019evolution}, who took a sample of 347 main sequence low mass stars observed by {\kep} and found flare activity decreases as they spin down with age. In addition, H$\alpha$ and X-ray emission is also saturated in rapid rotators where a decline is observed towards slower rotation periods \citep[e.g.][]{newton2016rotation, yang2017flaring}. 

In \citet{doyle2019}, we conducted a statistical analysis of the stellar flares from 149 low mass dwarfs (M0-M6V) using 2-min cadence lightcurves from {\tess} data covering Sectors 1--3. We identified nine low mass ultra fast rotating stars (UFRs) which have rotation periods, $P_{\rm rot}$\ltae0.3~d, and show low levels of flaring activity in their {\tess} lightcurves. No evidence was found that the lack of activity is related to age or rotational velocities. Given the rotation-activity relation \citep{hartmann1987rotation}, faster rotating stars should display higher levels of activity, so this comes as a surprise. In a further investigation, \citet{ramsay2020ufr} utilised {\tess} 2-min cadence data from Sectors 1--13 to identify more than 600 low mass (K9--M5V) stars with $P_{\rm rot}<$1~d. They found the fraction of stars showing flares drops significantly at $P_{\rm rot}<$0.2~d, compared to stars with rotation periods $0.6< P_{\rm rot}< 1.0$~d. The question remains, why do these rapidly rotating stars show little or no flaring activity? 

One possible explanation could be related to the magnetic field strength and configuration (i.e. the geometry of the magnetic field) of those stars. For example, \citet{kochukhov2017global} investigate the global and small scale magnetic field of the nearby M dwarf binary GJ 65~AB. These stars have nearly identical masses and rotation rates but have very different magnetic field configurations and strengths. As a consequence, they both display varying degrees of magnetic activity. The binary components of GJ 65~AB are both fully convective (they have spectral types M5.5-M6) and therefore must generate their magnetic field in a different way to solar-type stars which is driven in the tachocline, a boundary layer between the radiative and convective zones. Overall, the configuration and field strength of the magnetic field, along with the rotation period and age all play an important role in stellar magnetic activity. 

In this paper, we utilise FORS/VLT spectro-polarimetric data to search for the presence of a magnetic field in a sample of UFR low mass stars which have a spectral type in the range M2--M6V. We emphasise that this is a pilot study where we have selected bright targets within a range of rotation periods; our aim is to determine whether they have measurable magnetic fields and what implications this has on the wider rotation-activity relation. If they have a significant magnetic field then why do they not show more flares? Consequently, if they do not show a magnetic field then why not, given we expect the rapid rotation to drive a strong magnetic field through the stars dynamo mechanism? Why is this in contradiction to the rotation-activity paradigm which has been long established and is well accepted in stellar physics? 

In addition, we examine further {\tess} lightcurves obtained in Cycle 3 to investigate the long term variability of the stars in our sample. This includes identifying and determining the energy of the flares in both Cycles 1 and 3, establishing if the flare rate of our targets has changed over three years. We also explore whether there has been any change in the rotational modulation of the stars between the {\tess} Cycles and if so, what are implications? There is wide importance in these investigations of the magnetic field and flaring activity of these UFRs as it is directly related to many properties such as dynamo generation, magnetic activity and age of stars which are believed to be well understood. In a companion paper \citep{ramsay2021ufr}, we explore whether stars identified as UFRs could be binary stars which could help explain the lack of flares in {\tess} data of stars which otherwise appear to be low mass stars.

\begin{figure}
  \begin{center}
  \includegraphics[width=0.47\textwidth]{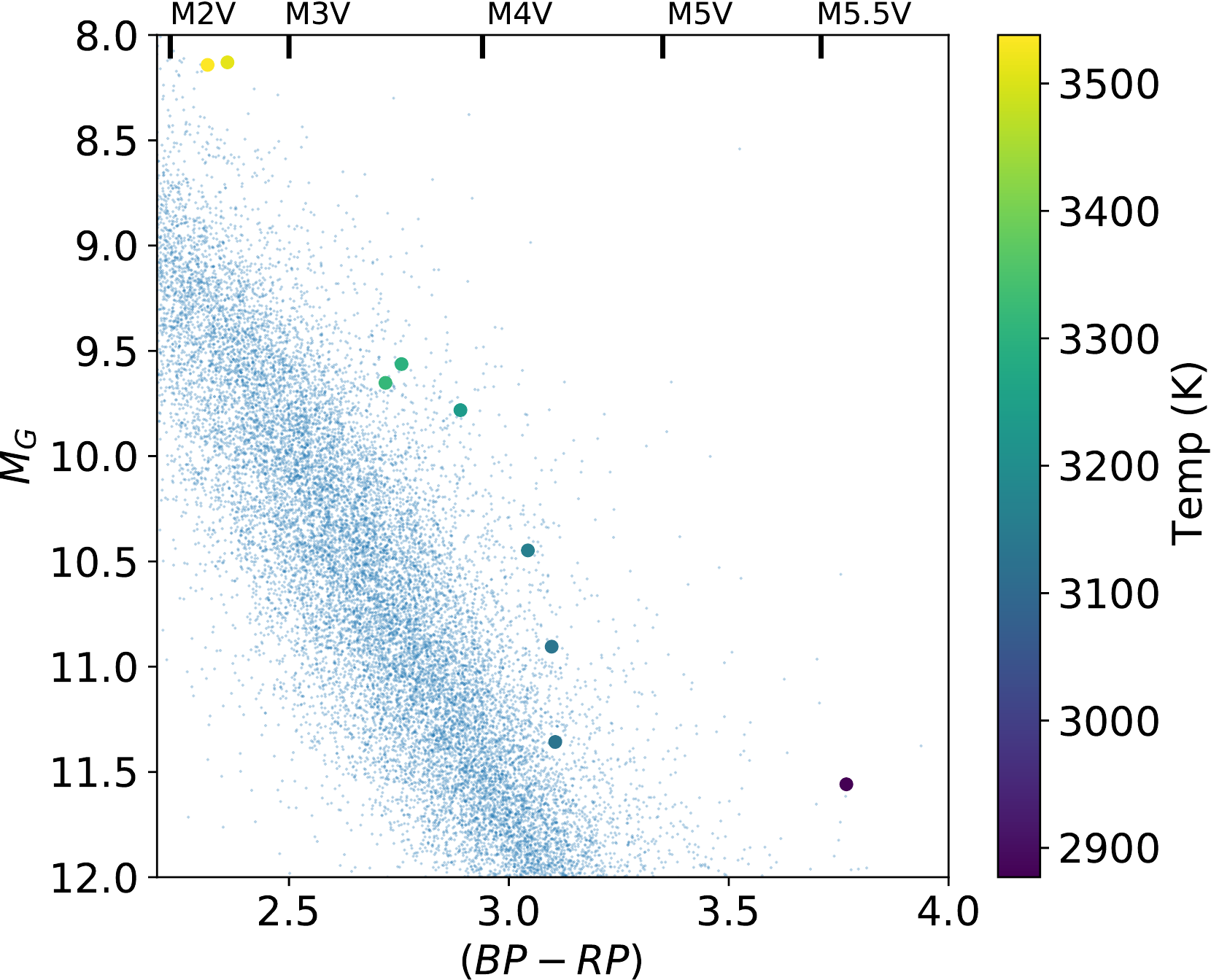}  
  \caption{The Gaia H-R Diagram where the small pale blue dots are stars within 50~pc of the Sun in the Gaia $(BP-RP)$, $M_{G}$ colour-absolute magnitude diagram \citep{gaia16, gaia18}. The larger coloured circles are the sources in our sample of UFRs, with the colour indicating the temperature of the star taken from the TIC \citep{stassun2018tess}. The colour of spectral sub-types has used the work of \citet{PecautMamajek2013}.}
    \label{gaiaHRD}
    \end{center}
\end{figure}

\section{Low Mass Star Sample}

In selecting low mass stars to search for a magnetic field, we initially selected those which showed a modulation on a period $<$1~d in the study of \citet{doyle2019}. To confirm the rotation period of our sample, we used the generalised Lomb Scargle (LS, \citet{Zechmeister2009,NRF1992}) and Analysis of Variance (AoV, \citet{Schwarzenberg-Czerny1996}) periodograms to identify the most prominent period in each of the stars light curves from each sector. Apart from TIC 425937691 (where we find a period of 0.201 d -- twice that originally identified in \citet{doyle2019}) we find that the LS and AoV periodograms produce periods which are consistent with each other and with \citet{doyle2019}. For the purposes of this work, we take the most prominent period to be the rotation period of the star. We then identified ten stars in the period range 0.0--1.0~d which had five equal bins each with two stars -- one showing relatively few flares with {\tess} and one showing more. In \citet{ramsay2020ufr} we define flare inactive stars to be those with flare rates $<$0.044 flares per day (corresponding to at least one flare in the shortest observation length of the sample). Therefore, compared to the sample of \citet{ramsay2020ufr}, all stars in the sample reported in this paper are considered `flare active'. Our final constraint was that each star had to be bright enough so that one polarised spectrum was obtained in a time shorter than 0.2 rotation cycles. This ensured any line or magnetic feature was not blurred due to the rapid rotation. All of our targets are brighter than $T_{mag} = 13.2$ (the stars brightness in the {\tess} band, \citet{Sullivan2015}) with spectral types between M2 and M5.5: details of each star are shown in Table \ref{stellar_properties}. Although previous studies  \citep[e.g.][and references therein]{wright2011stellar} have indicated that stars with spectral types later than M3--M4V are fully convective, recent work by \citet{mullan2020transition} indicates that the dynamo transition to fully convective stars can be narrowed down to between M2.1 and M2.3. It is, therefore, likely that most of our targets are fully convective. 

We show the location of our targets on the Gaia HRD in Figure \ref{gaiaHRD}. Our targets appear redder or brighter than the low mass main sequence track. This could be a result of youth, leading to radius inflation and temperature suppression \citep[e.g.][]{Stassun2012,SomersStassun2017}, or it could be due to the star being a member of a binary. We were able to estimate ages for seven out of the ten stars in our sample by their association with moving groups or associations (see the caption of Table \ref{stellar_properties} for references). As a result, we have ages between 13 -- 150~Myr, indicating they are young stars.

Two of the stars in our sample, UCAC3 53-724 (TIC 425937691) and UPM~J0113-5939 (TIC 206544316), were also part of the study of \citet{zhan2019complex} who searched {\tess} 2-min cadence lightcurves in Sectors 1 and 2 for highly structured rotational modulation patterns in low mass stars. These lightcurves display complex, highly structured periodic variations which cannot be explained using simple starspot models. They conclude the most likely scenario to explain the highly structured nature of the rapidly rotating M dwarfs to be a spotted host star rotating obliquely inside of a dusty ring of material a few tens of stellar radii away. Determining whether these stars have a detectable magnetic field will help test models of these stars.

To indicate if any of our stars are members of binary systems we used the Renormalised Unit Weight Error (RUWE) from the Gaia EDR3 Catalog \citep{gaiaedr3brown}. This is a goodness-of-fit measurement of the single star model to the targets astrometry which is highly sensitive to the photocentre motion of binaries themselves. Overall, a RUWE $>$ 1.4 could indicate the star is non-single, however, RUWE values between 1.0 -- 1.4 may also signify binaries \citep{stassun2021parallax}. In our sample, nine of our targets were present in ERD3 with three having a RUWE $>$ 2. A further three targets had RUWE between 1.4 and 1.6 with the remaining three having RUWE $\sim$1.2, see Table \ref{stellar_properties}. Overall, this indicates that at least three of our targets (TIC 425937691, TIC 229142295 and TIC 220539110) are possible members of binary systems. In \S \ref{rvsearch} we search for radial velocity variations between the epochs of the FORS2/VLT data.

\section{Observations}
\subsection{FORS2/VLT}
\label{Sect_LS}

To search for evidence of magnetic fields in our sample of ten low mass UFRs, we used the FORS2 instrument \citep{appenzeller1992fors, appenzeller1998fors} of one of the ESO 8.2 m VLT to obtain circular spectropolarimetry. FORS2 is a multipurpose instrument capable of imaging and low resolution spectroscopy equipped with polarimetric optics, consisting of a retarder waveplate (either a $\lambda/2$ for linear polarisation or a $\lambda/4$ for circular polarisation), and a Wollaston prism. For our observations we used the beam-swapping technique \citep[e.g.][]{Bagetal09}, setting the $\lambda/4$ waveplate at position angles $-45 +45 +45 -45\degr$. All stars were observed both with grism 1200R and order separating filter OG435 and grism 1028z with order separating filter OG590, using a 1\arcsec\ slit width. Grism 1200R cover the spectral range $5750-7310$\,\AA\ with a spectral resolution of $\sim 2140$; grism 1028z cover the spectral range $7730-9480$\,\AA\ with a spectral resolution of 
$\sim 2560$. 

Data were reduced as outlined in \citet{BagLan18}. In particular the mean longitudinal magnetic field \bz\ (the component of the magnetic field along the line of sight, averaged over the visible stellar disk) was measured using the relationship:

\begin{equation}
\frac{V}{I} = - g_\mathrm{eff} \ \cz \ \lambda^{2} \
                \frac{1}{I} \
                \frac{\mathrm{d}I}{\mathrm{d}\lambda} \
                \bz\;,
\label{Eq_Bz}
\end{equation}
where $I$ and $V$ are the Stokes parameters describing the intensity and the circular polarisation, respectively, $g_\mathrm{eff}$ is the effective Land\'{e} factor, and

\begin{equation}
\cz = \frac{e}{4 \pi m_\mathrm{e} c^2}
\ \ \ \ \ (\simeq 4.67 \times 10^{-13}\,\mathrm{\AA}^{-1}\ \mathrm{G}^{-1}) \; ,
\end{equation}
where $e$ is the electron charge, $m_\mathrm{e}$ the electron mass and
$c$ the speed of light. To estimate \bz\ we used a least-squares technique and minimised the
expression:

\begin{equation}
\chi^2 = \sum_i \frac{(y_i - \bz\,x_i - b)^2}{\sigma^2_i}\; ,
\label{Eq_LS}
\end{equation}
where, for each spectral point $i$, $y_i = V(\lambda_i)/I(\lambda_i)$, $x_i = -g_\mathrm{eff} \cz \lambda^2_i (1/I_i\ \times \mathrm{d}I/\mathrm{d}\lambda)_i$, and $b$ is a constant introduced to account for possible spurious polarisation in the continuum. 

In addition to the reduced Stokes parameter $V/I$, we have also calculated the so-called null profile $N$ \citep[e.g.][]{Bagetal09}. The $N$ profile is expected to oscillate about zero within the same uncertainties as those of $V/I$, and, as such, allows us to perform a quality check on the data. In particular it is possible to calculate the null field \nnz\ using Eqs.~(\ref{Eq_Bz}) and (\ref{Eq_LS}) after replacing $V$ with $N$, and check that the results are consistent with zero. 

\begin{figure*}
   \centering
   \includegraphics*[width=12cm,angle=270,trim={0.9cm 0.0cm 0.7cm 0.0cm},clip]{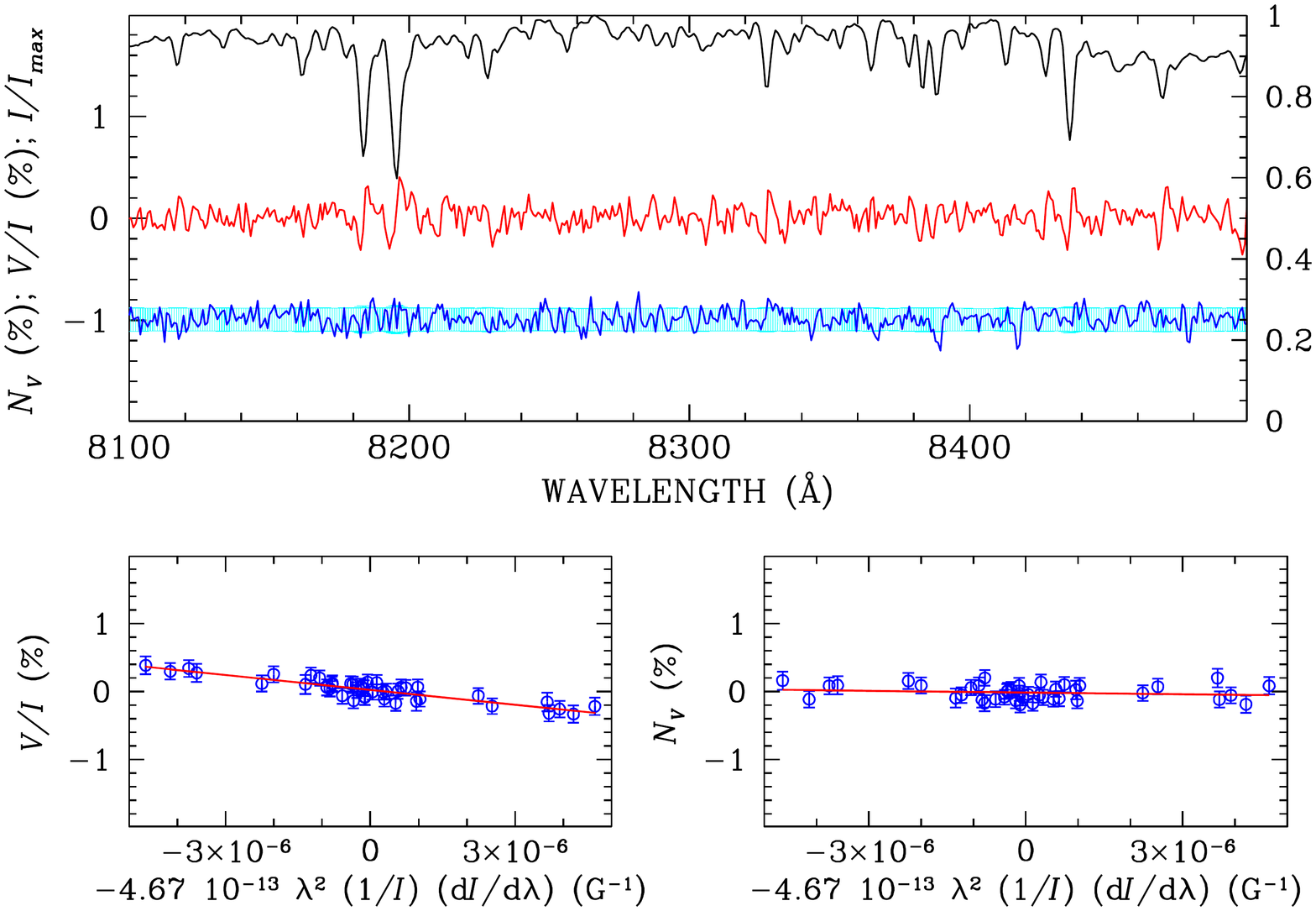}
      \caption{Spectro-polarimetric observations of EXO 0235.2-5216 obtained on 2020-11-20. {\it Top panel}: observed flux in arbitrary units (black solid line),  $V/I$ (red line) and $N/I$ (blue line) spectra obtained with grism 1028z. The null profile is offset by $-1$\,\% for display purposes and is superposed to photon-noise error bars of $V/I$ which appear as a light blue background. {\it Bottom panels}: $V/I$ (left panel) and $N_V$ (right panel) versus  $ - g_\mathrm{eff} \ \cz \ \lambda^{2} \
                \frac{1}{I} \
                \frac{\mathrm{d}I}{\mathrm{d}\lambda} $, as explained in the text.
              }
     \label{Fig_EXO}
\end{figure*}

The spectra of M dwarfs can show regions which are dominated by heavily blended atomic lines, or by molecular lines, some of which may have negative Land\'{e} factors. We therefore apply Eq.~(\ref{Eq_Bz}) to the H$\alpha$ emission line (probing the chromosphere) and from the Na~I lines at 8183 and 8195\,\AA\ (probing the photosphere), with the results being reported in Table~\ref{Tab_Log}. Figure~\ref{Fig_EXO} shows the observed spectrum of EXO 0235.2-5216 where, in the top panel, we show the Stokes $I$ and $V/I$ profiles (as well as the null profile) and the bottom panels show the best-fit obtained by minimising the $\chi^2$ expression of Eq.~(\ref{Eq_LS}) using the $V/I$ profile (left panels) and the null profile (right panel) around the Na~I lines.

\begin{table*}
\caption{The flare properties of the ten low mass stars in our survey. This includes the flare rates from Cycle 1 \& Cycle 3; the total flare number; duration range; energy of the flares and the average flare energy (i.e. the sum of the total flare energies for each object divided by the total monitoring time). We also include the rotation period of each star which was obtained from \citet{doyle2019}, the stellar spectral types ({from \tt SIMBAD}) and stellar masses and radii from the TIC \citep{stassun2018tess}. The stars have been split into their period bins using horizontal lines and stars with a detectable magnetic field have been highlighted in bold. Note: $^*$ these targets have been identified as complex rotators in \citet{zhan2019complex}.}
   \begin{center}
   \label{flare_properties}
\resizebox{1.0\textwidth}{!}{
	\begin{tabular}{lcccc|ccc|ccc|cc}
	\hline
	   &    & & & & \multicolumn{3}{|c|}{Cycle 1} & \multicolumn{3}{|c|}{Cycle 3} \\
    \cline{6-11}

	TIC ID    & $P_{\rm rot}$ & SpT & Mass  & Radius & Flare Rate & $log(E)$ & Duration Range & Flare Rate  & $log(E)$      & Duration Range & Obs. Length & Avg. Flare Energy \\
	          &   (d)     &     & \Msun & \Rsun & (per day)   & (erg)    & (min)          & (per day)   & (erg)         & (min)          & (days)             & ($log(E)$, erg)   \\
	\hline
    {\bf 158596311} & 0.154   & 4.1   & 0.36 & 0.37 &  0.16        &  32.06 -- 32.77  & 5.8 -- 25.9    & 0.16        & 31.89 -- 33.38 & 5.8 -- 36.0    &  44.08              & 26.98              \\
    425937691$^*$ &	0.201   & 5.5   & 0.31 & 0.32 &  0.65        &  31.52 -- 33.42  & 5.8 -- 40.3    & 0.30        & 31.80 -- 34.23 & 5.8 -- 70.6    &  69.96             & 27.65              \\
    \hline
    {\bf 206544316}$^*$ &	0.322  & 3.7    & 0.46 & 0.46 &  0.39        &  32.55 -- 34.93  & 5.8 -- 99.4    & 0.42        & 32.16 -- 34.74 & 5.8 -- 139.7   &  92.45             & 28.39              \\
    156002545 &	0.353  & 3.4    & 0.45 & 0.46 &  0.12        &  32.99 -- 33.85  & 21.6 -- 90.7   & 0.09        & 32.26 -- 32.79 & 5.8 -- 17.3    &  45.30             & 27.42              \\
    \hline
    {\bf 248354845} &	0.522  & 3.5    & 0.47 & 0.48 &  0.86        &  32.00 -- 34.39  & 5.8 -- 119.5   & 0.90        & 31.84 -- 33.73 & 5.8 -- 61.9    &  40.67              & 28.06              \\
    229142295 &	0.447  & 4.5    & 0.24 & 0.27 &  0.09        &  31.47 -- 32.61  & 5.8 -- 40.3    & 0.13        & 31.31 -- 32.29 & 5.8 -- 25.9    &  87.78             & 25.98              \\
    \hline
    220539110 &	0.773   & 3.0   & 0.65 & 0.68 &  0.53        &  32.17 -- 34.56  & 5.8 -- 63.4    & 0.44        & 32.10 -- 33.52 & 5.8 -- 34.6    &  133.42             & 27.82              \\
    {\bf 166808151} & 0.740  & 2.0    & 0.66 & 0.69 &  1.14        &  32.05 -- 34.56  & 5.8 -- 90.7    & 0.95        & 32.07 -- 34.17 & 5.8 -- 47.5    &  88.64             & 28.24              \\
    \hline
    {\bf 141807839} &	0.839  & 4.5    & 0.61 & 0.63 &  0.58        &  32.18 -- 34.81  & 5.8 -- 277.9   & 0.71        & 32.19 -- 34.48 & 5.8 -- 112.32  &  465.24           & 28.08              \\
    201861769 & 0.862   & 4.1   & 0.30 & 0.32 &  0.07        &  32.58 -- 33.26  & 11.5 -- 34.56  & 0.18        & 31.91 -- 32.50 & 5.8 -- 15.8    &  46.51              & 26.80              \\
    \hline
    \end{tabular}}
    \end{center}
\end{table*}

\subsection{TESS Photometric Lightcurves}

The Transiting Exoplanet Survey Satellite \citep[{\tess}:][]{Ricker2015} was launched in April 2018 with a primary mission of searching for exoplanets via the transit method around low mass M dwarf stars. In its first two years, {\tess} has completed a near all-sky survey observing more than 200,000 of the closest stars to our Sun with a cadence of 2-min. In \cite{doyle2019}, we analysed the flare properties of the ten stars in this sample utilising {\tess} lightcurves of Cycle 1 from Sectors 1--3. In this study, we will use lightcurves from both Cycles 1 and 3 to compute a similar analysis, while also investigating any change in their flaring activity. 

Photometric lightcurves used in this analysis were made between 25th July 2018 - 18th July 2019 for Sectors 1--13 and 4th July 2020 - 26th May 2021 for Sectors 27 - 38. At the time of writing, only data up to Sector 38 was available, however, this includes all data for our targets in Cycle 3. We downloaded the calibrated lightcurves for each of our target stars from the MAST data archive\footnote{\url{https://archive.stsci.edu/tess/}}, using the data values for {\tt PDCSAP\_FLUX}. All points which did not have {\tt QUALITY=0} were removed and each lightcurve was normalised by dividing the flux of each point by the mean flux of the star \citep[see][for more details]{doyle2019,ramsay2020ufr}.

The size of the {\tess} pixels (21$^{''}$/pixel) are sufficiently large that dilution of the target star by spatially nearby stars could occur. In addition, if a nearby star was variable, the signature of this variation could be present in the light curve of the target star. Although the FWHM of the {\tess} PSF is 1.9 pixels, the number of pixels which are used to extract the lightcurves in the {\tess} pipeline is typically 3--4.  We therefore searched for all stars within 1$^{'}$ of our targets using {\tt tpfplotter} \citep{Aller2020}. There were only two target stars which had another star within 1$^{'}$ which were at most 2 mag fainter than the target: TIC 141807839 had a star 2.0 mag fainter and 36.6$^{''}$ distant, and TIC 206544316 had a star 1.2 mag fainter and 44.5$^{''}$ distant. Neither were in the aperture of the target star nor were they stars with spectral type M0V or later. Therefore, we conclude that the {\tess} light curves were not significantly affected by blending issues.

\subsubsection{Stellar Flare Identification}

We identified flares present in the {\tess} lightcurves for each star on a sector by sector basis. To do this, we implemented the open source {\tt Python} software {\tt AltaiPony} \citep{ilin2021flares} to automatically detect and characterise flares in our sample. Before each lightcurve could be input into {\tt AltaiPony}, it had to be flattened with rotational modulation trends being removed. As a result, each lightcurve was fitted with a Savitzky-Golay filter to detrend the lightcurve. However, for two of our targets, UCAC3~53-724 (TIC~425937691) and UPM~J0113-5939 (TIC~206544316), this was insufficient to remove the periodic signal and so a custom detrending\footnote{\url{https://github.com/ekaterinailin/TESS_UCD_flares}} was utilised. This custom detrending had a few other steps including removing global trends, fitting a Lomb-Scargle periodogram to remove strong rotational modulation and masking and padding outliers. 

The flare classification is based on \citet{davenport2014kepler} and \citet{davenport2016kepler}, where flares are identified as two or more consecutive points which lie 2.5$\sigma$ higher than the mean \citep[see][for more details]{doyle2018investigating, doyle2019}. {\tt AltaiPony} determines many flare properties including the start and stop times, flare amplitude and equivalent duration. We used the start and stop times to calculate the duration of the flares and the equivalent duration was multiplied by the quiescent luminosity (see Table \ref{stellar_properties}) to determine the flare energy in the {\tess} bandpass. The quiescent luminosity was determined in \citet{doyle2019} by converting multi-colour magnitudes to flux and fitting them using a polynomial to produce a template spectrum, which was convolved with the TESS band-pass to derive the stars quiescent flux. Gaia parallaxes were then inverted to provide distances to each star which was used to determine the quiescent stellar luminosity. Overall, we identified 605 flares across the ten low mass stars in our sample with energies between 2.0$\times$10$^{31}$ and 5.4$\times$10$^{34}$~erg. Full details of the flare properties for each star are listed in Table \ref{flare_properties}. 

\section{Results}

Here we discuss the independent results from the ESO VLT/FORS2 and {\tess} data, before bringing them together in \S \ref{discussion}. 

\begin{table*}
\caption{\label{Tab_Log} Observing log and field measurements from the VLT/FORS2 spectropolarimeter of the ten UFR low mass stars in our sample. Stars with a detectable magnetic field have been highlighted in bold.}
\tabcolsep=0.14cm
\begin{tabular}{ ll c lrr r@{$\pm$}l r@{$\pm$}l c crr r@{$\pm$}l r@{$\pm$}l } 
\hline\hline
\multicolumn{2}{c}{}                     &  
$\phantom{xx}$&
\multicolumn{7}{c}{GRISM 1200R (H$\alpha$)}      &  
$\phantom{xx}$&
\multicolumn{7}{c}{GRISM 1028z (Na)}     \\  
\multicolumn{1}{c}{STAR}                 & 
\multicolumn{1}{c}{DATE}                 & 
$\phantom{xx}$&
\multicolumn{1}{c}{UT}                   & 
\multicolumn{1}{c}{EXP}                  & 
\multicolumn{1}{c}{{\it S/N}}            & 
\multicolumn{2}{c}{$\langle B_z \rangle$}& 
\multicolumn{2}{c}{$\langle N_z \rangle$}& 
$\phantom{xx}$&
\multicolumn{1}{c}{UT}                   & 
\multicolumn{1}{c}{EXP}                  & 
\multicolumn{1}{c}{{\it S/N}}            & 
\multicolumn{2}{c}{$\langle B_z \rangle$}& 
\multicolumn{2}{c}{$\langle N_z \rangle$}\\ 
\multicolumn{1}{c}{    }                 &  
\multicolumn{1}{c}{yyyy-mm-dd}           &  
$\phantom{xx}$&
\multicolumn{1}{c}{hh:mm}                 &  
\multicolumn{1}{c}{(s)}                  &  
\multicolumn{1}{c}{\AA$^{-1}$}           &   
\multicolumn{2}{c}{(G)                  }&  
\multicolumn{2}{c}{(G)                  }&  
$\phantom{xx}$&
\multicolumn{1}{c}{hh:mm}                   &  
\multicolumn{1}{c}{(s)}                  &  
\multicolumn{1}{c}{\AA$^{-1}$}            &  
\multicolumn{2}{c}{(G)                  }&  
\multicolumn{2}{c}{(G)                  }\\  
\hline
{\bf AL 442}&2020-11-06 &&07:13 &1200 &780 &$  -80 $&105&$ 210$& 55&&  07:33&  640& 805&$  -580$& 95&$ 155$& 75\\
{\bf (TIC 141807839)}&2020-11-25 &&05:17 &1200 &670 &$ -160 $& 55&$ -50$& 35&&  05:38&  640& 790&$  -580$&100&$  80$& 80\\ 
            &2020-11-26 &&03:25 &1200 &710 &$   35 $& 75&$ -10$& 60&&  03:45&  640& 825&$  -590$& 95&$ 165$& 75\\
            &2020-11-28 &&04:09 &1200 &710 &$ -230 $& 45&$  45$& 20&&  04:25&  320& 590&$  -885$&140&$  30$&110\\[3mm]

UCAC3 53-724    &2020-11-06 &&01:05 &1200 &205 &$ -150 $&230&$ 220$&105&&  01:28& 1200& 440&$    75$&175&$ 330$&185\\
(TIC 425937691)&2020-11-13 &&01:54 &1200 &250 &$ -277 $&355&$ 210$&130&&  02:18& 1200& 515&$   355$&150&$ -80$&130\\[3mm]

{\bf UCAC4 204-001345}&2020-11-06&&04:41&1200&370 &$  -10 $&205&$-135$&265&&  05:07& 1360& 615&$   215$&185&$-110$&195 \\
{\bf (TIC 158596311)}&2020-11-06 &&05:34 &1200 &440 &$  170 $&185&$-380$&120&&  06:01& 1360& 755&$   130$&145&$ -85$&130 \\
            &2020-11-24 &&04:01 &1200 &445 &$   85 $&135&$ 190$&125&&  04:27& 1360& 800&$   415$&115&$ 125$&100 \\
            &2020-11-26 &&04:14 &1200 &435 &$  205 $&190&$-120$& 75&&  04:40& 1360& 760&$   575$&190&$ 260$&130\\
            &2020-11-27 &&02:12 &1200 &455 &$  215 $&160&$-355$& 80&&  02:39& 1360& 800&$   755$&150&$-295$&105 \\[3mm]

2MASS J0033-5116 &2020-11-24 &&01:14 &2400 &965 &$  -75 $& 75&$  75$& 50&&  01:15&  880& 810&$  -205$&115&$  85$&115 \\
(TIC 156002545) &2020-11-27 &&01:17 &1200 &715 &$   40 $& 50&$  50$&105&&  01:41&  880& 920&$  -235$&115&$ 165$& 95 \\[3mm]

2MASS J0146-5339 &2020-11-24 &&02:11 &1200 &730 &$ -195 $&120&$  35$& 90&&  02:34&  960&1080&$    -5$&65&$ -20$&60 \\
(TIC 229142295) &2020-11-26 &&05:14 &1200 &715 &$ -145 $&110&$-225$&110&&  05:37&  960&1035&$    15$&65&$ -15$&70 \\
            &2020-11-27 &&03:54 &1200 &735 &$ -100 $&105&$  95$& 85&&  04:17&  960&1105&$   -85$&55&$  25$&70 \\[3mm]

2MASS J0232-5746 &2020-11-24 &&03:06 &1200 &340 &$ -165 $&205&$  30$& 85&&  03:29& 1200& 555&$   -80$&35&$ -25$&35 \\
(TIC 201861769) &2020-11-27 &&03:13 &1200 &330 &$   35 $&125&$  35$& 95&&  03:30&  480& 345&$  -140$&70&$ 135$&70 \\[3mm]

GSC 08859-00633   &2020-11-06 &&06:33 & 560 &985 &$ -160 $&105&$ 220$& 70&&  06:47&  360&1100&$   160$&60&$  80$& 80 \\
(TIC 220539110) &2020-11-24 &&04:57 & 560 &050 &$  200 $&105&$ -45$&125&&  05:12&  360&1105&$    10$&70&$-210$& 80 \\
            &2020-11-26 &&06:08 & 560 &010 &$ -210 $& 85&$ 215$&105&&  06:22&  360&1075&$   265$&95&$ 115$& 65 \\
            &2020-11-27 &&04:51 & 560 &990 &$   90 $&145&$  90$&125&&  05:05&  360& 995&$   -45$&60&$-300$& 80 \\
            &2020-11-28 &&01:04 & 560 &940 &$  370 $& 95&$ 110$&125&&  01:18&  360& 975&$  -130$&80&$  90$&100 \\[3mm]

{\bf GSC 04683-02117}&2020-11-30&&01:01&1200&810 &$  -90 $& 65&$  20$& 80&&  01:23&  800& 980&$ -570$& 85&$ -30$& 80 \\
{\bf (TIC 248354845)} &2020-12-03 &&03:39 &1200 &805 &$  -10 $& 90&$  80$&120&&  04:01 &   800 & 905 &$ -460$&65 &$-45$&60               \\[3mm]


{\bf UPM J0113-5939}&2020-11-14 &&02:32 &1200 &670 &$  240 $& 90&$ -13$&115&&  02:57& 1200&1070&$ 690 $&115&$  75$&65 \\
{\bf (TIC 206544316)} &2020-11-23 &&05:11 &1200 &625 &$ -115 $& 90&$  -2$&110&&  05:36& 1200&1000&$ -50 $& 50&$ -10$&85 \\[3mm]

{\bf EXO 0235.2-5216}&2020-11-07 &&08:15 & 400 &875 &$ -335 $& 45&$ -15$& 40&&  08:26&  240& 910&$ -655 $&115&$ 120$&75 \\
{\bf (TIC 166808151)}  &2020-11-12 &&08:12 & 400 &950 &$  -45 $& 75&$   6$& 80&&  08:24&  240& 975&$ -725 $& 65&$ -80$&85 \\
\hline
\end{tabular}
\end{table*}

\subsection{Magnetic Fields with FORS2}

We report the results of applying Eq.~(\ref{Eq_Bz}) to the H$\alpha$ emission line and the Na\,I absorption lines at 8183 and 8195\,\AA\ for each of the spectra in Table~\ref{Tab_Log}. We now apply a Bayesian approach to determine the likelihood of whether each star has a detectable magnetic field and place limits of those which do not. Following \citet{PetitWade2012}, we define stars as having very strong evidence (an odd-ratio $>$100 \citep[see Eq. (3.14) of][]{gregory2005bayesian}); strong evidence (30--100); moderate evidence (10--30); weak evidence (3--10) or no evidence ($<$3) for having a magnetic field.

The Stokes V profile is sensitive to the component of the local magnetic field along the line of sight. Each element of the visible hemisphere of a star is characterised by a different longitudinal component of the local magnetic field, but also by its own radial component of the rotational velocity. Because of this differential Doppler effect, even if the longitudinal field integrated over the stellar disk is zero, the observed Stokes V profile may have a non-zero amplitude (if a field is present at the surface of the star). However, the spectral resolution of our observations is too low to detect this subtle effect. The mean longitudinal field is non-zero only for a morphology that varies smoothly over the stellar surface, like that of a dipolar field. It is, therefore, practical to interpret our measurements as a constraint on the dipolar component of the field, keeping in mind that with low-resolution spectropolarimetry, a tangled magnetic field would pass undetected.

Indeed, even a strong dipolar field could be undetected if, at the time of the observations its axis is perpendicular to the line of sight. However, since the stars have been repeatedly observed, this gives a stronger constraint because it is unlikely that a dipolar field appears every time we observe it perpendicular (or almost perpendicular) to the line of sight. Because of stellar rotation, the dipolar axis is likely viewed under different angles at different epochs, unless the repeated observations are short compared to the rotation period. Similarly, repeated non-zero field measurements can be used to estimate a range of field strengths that the star is likely to possess under the assumption that the field has a dipolar morphology (we have to note the probability density function of the tilt angle or the rotation axis with respect to the line of sight, and the angle between the dipolar axis and the rotation axis). Our data come with the additional constraint of the stellar rotation period, known from the {\tess} photometry.

Even under the assumption of a dipolar morphology, determining quantitative answers to our questions is not trivial but has been tackled using a Bayesian approach by \citet{KolenbergBagnulo2009,PetitWade2012,AsensioRamos2014,Bagnulo2020}. Specifically we calculate the `odd-ratio' \citep[see Eq. (3.14) of][]{gregory2005bayesian} and use the probability density function of the field strength of the dipolar field as given by Eq. (2) of \citet{Bagnulo2020}. This is modified to take into account the constraint given from the fact that we know the stars rotational period.

To summarise, we find strong evidence for a magnetic field in AL~442, UCAC4 204-0011345, GSC 04683-02117, EXO 0235.2-5216 and UPM J0113-5939. Of these five, four were the more flare active stars within the period bins with one being the less flare active star. We find weak evidence for a field in GSC 08859-00633 and no evidence for a field in the remaining stars in our sample (one of these is the second fastest rotating star UCAC3 53-724 which has a rotation period of 0.2~d: we discuss this further in \S \ref{discussion}). The typical field strength (expressed as mean field modulus over the stellar surface) is of order 1--2~kG for the magnetic stars, while for the non magnetic stars there is a 95 percent probability that the field is weaker than $\sim$2~kG (if present at all). 2MASS J0146-5339 is an exception to this as our observations set the upper limit of the surface field to a few 100~G. We note that any field measurements of a binary system will be diluted as the observations will refer to two stars, therefore, it is difficult to say which or if both possess a magnetic field. We go on to discuss the issue of binarity in further detail in Section \ref{rvsearch}.

\begin{figure}
  \begin{center}
  \includegraphics[width=0.47\textwidth]{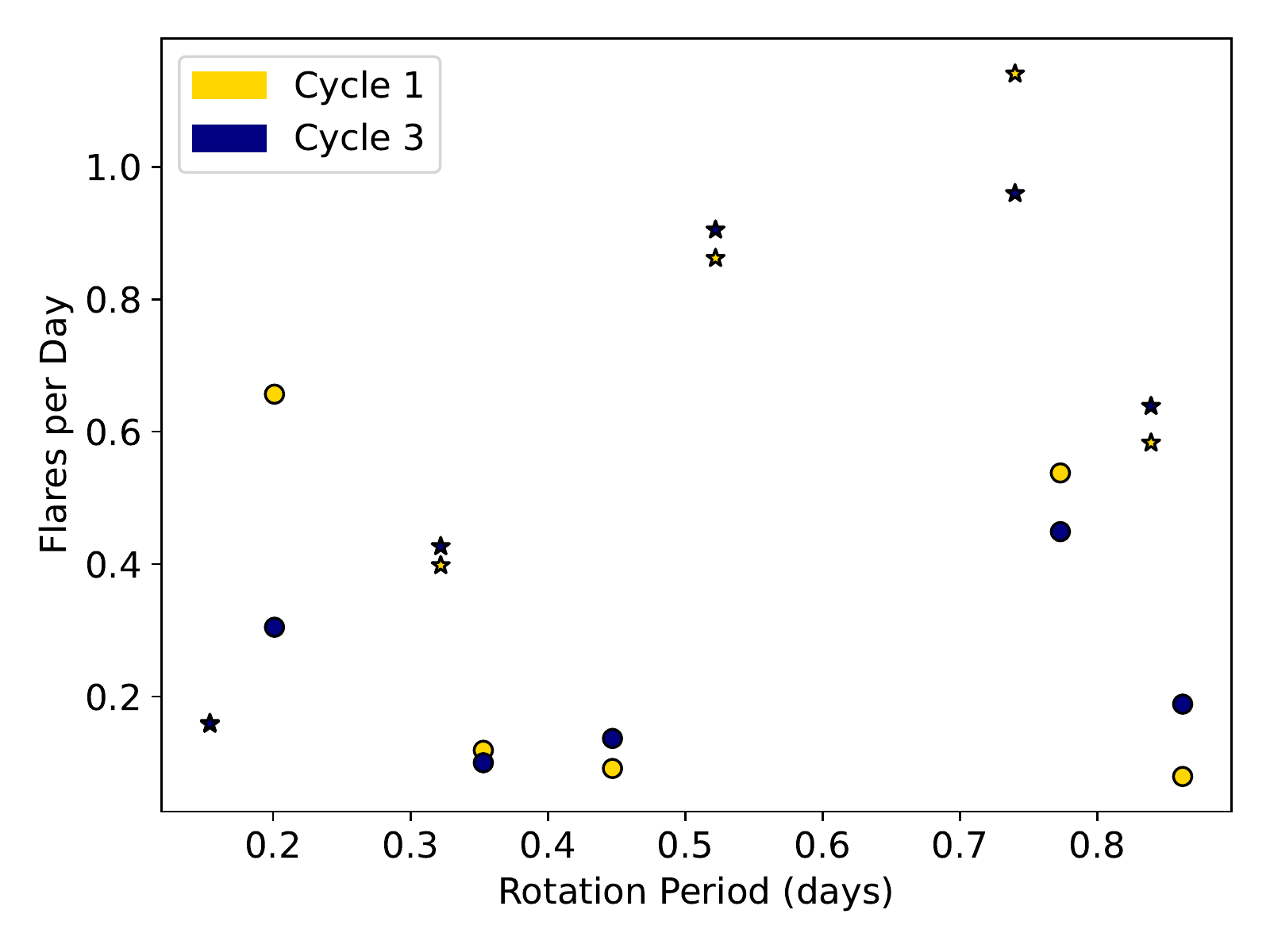}     
  \caption{The flare number per day as a function of rotation period for both Cycles 1 and 3. The star markers indicate targets where a magnetic field was detected and the last three digits of the TIC ID for each star is highlighted. (Note: the star with the shortest rotation period has an identical flare rate in each Cycle).}
    \label{compare_cycles}
    \end{center}
\end{figure}

\subsection{Radial Velocity Search with FORS2}
\label{rvsearch}

To check for any evidence of binarity amongst our sample of targets, we use the FORS2 spectra to search for radial velocity (RV) variations. We would like to note that each observing series at each epoch consists typically of eight exposures taken within minutes from each other. For each star we consider the very first exposure as a reference spectrum. Each subsequent spectrum is then convolved with the reference spectrum to calculate the RV shift for both stellar lines and telluric lines (telluric lines are taken as absolute reference to correct for instrument flexures). We used only the spectra obtained with grism 1028z in which one pixel corresponds to 0.84~\AA. 
Overall, our analysis reveals clear RV modulation in GSC~08859-00633 and line splitting in some of the spectra of GSC~04683-02117. Further analysis of the GSC 08859-00633 spectra revealed a K1 velocity of 19.45 $\pm$ 0.86 km/sec, assuming a purely sinusoidal modulation (zero eccentricity) and a 0.773~d period from {\tess}. This yields a mass function (minimum mass) $f$ = 0.00059~{\Msun} for the unseen companion object. Assuming a random inclination, we find that in 50\% of cases the mass of the companion is < 55~$M_{Jup}$ and in 90\% of cases < 110~$M_{Jup}$. The source is thus likely a M dwarf - brown dwarf binary. The remaining sources do not show any clear changes in their RVs and show an RMS scatter of RV values in range 1.0 to 7.5 kms$^{-1}$.

\subsection{Long-term Variability with {\tess}}

Our sample of M dwarfs have been observed in both Cycles 1 and 3 in 2-min cadence mode, providing an opportunity to compare the magnetic variability (i.e. stellar flares and starspots) of these stars up to three years apart. In Figure \ref{compare_cycles}, we show the flare number per day for each star in both Cycles 1 and 3 and as a function of rotation period. In terms of rotation, there is very little change in flaring activity of stars with $P_{\rm rot} <$ 0.6~d (TIC~425937691 is an exception to this). The remaining stars with $P_{\rm rot} >$ 0.6~d show a change of between $\sim$0.1 and 0.2 flares per day. To determine whether the differences observed for each star were significant, we simulated results assuming Poisson statistics. For each star, the rate from the cycle which shows the higher rate was used and the expected number of flares was simulated for the lower rate cycle (this was done 10$^{6}$ times). The fraction of simulated rates which produces the observed number of flares (or less) was counted in the lower rate cycle. Overall, this analysis showed there are no significant differences in the flare rates in any of the stars from our sample between the two cycles, even at a 3 $\sigma$ level. 

\begin{figure}
  \begin{center}
  \subfloat{\includegraphics[width=0.49\textwidth]{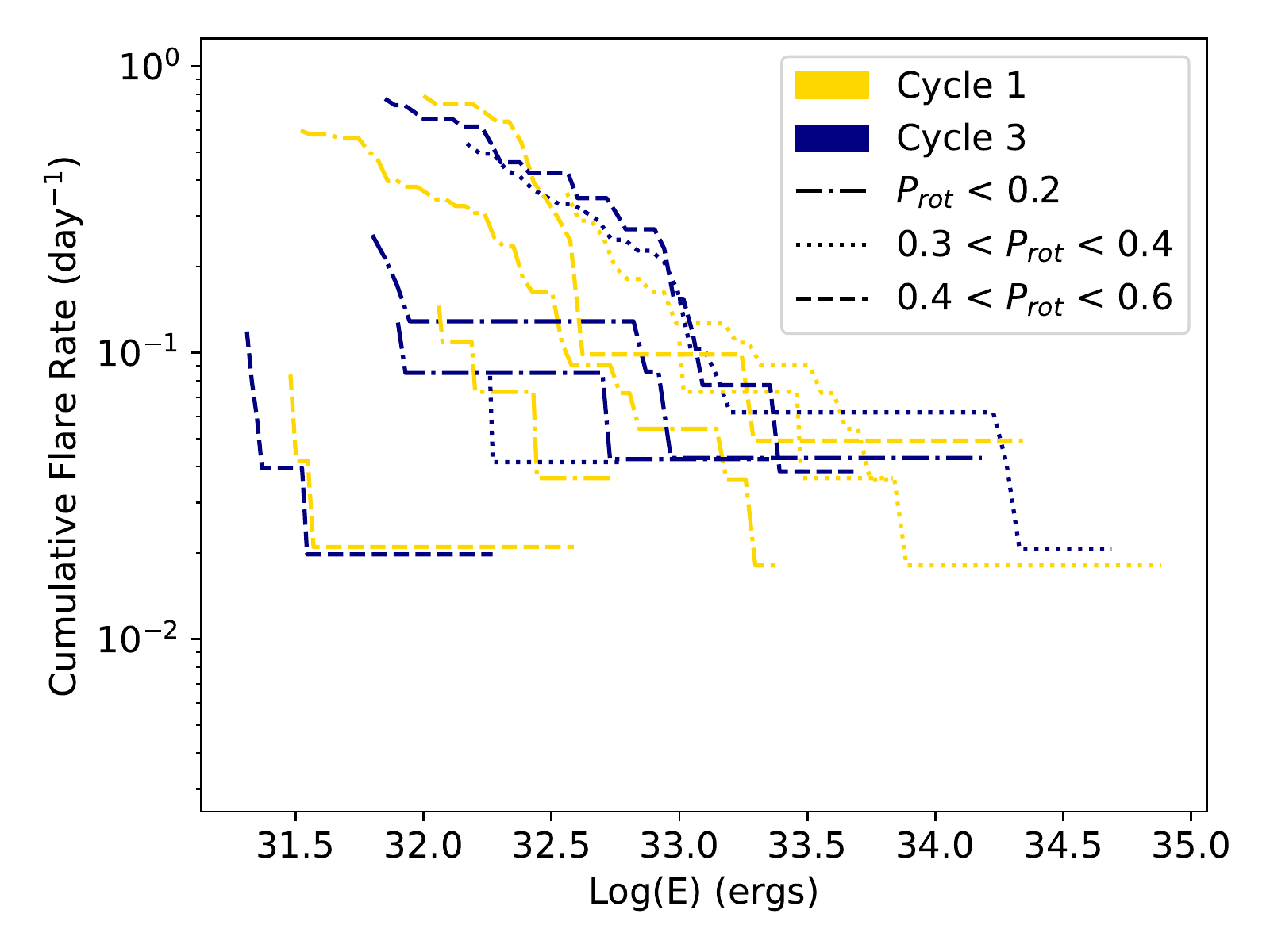} }\\
  \subfloat{\includegraphics[width=0.49\textwidth]{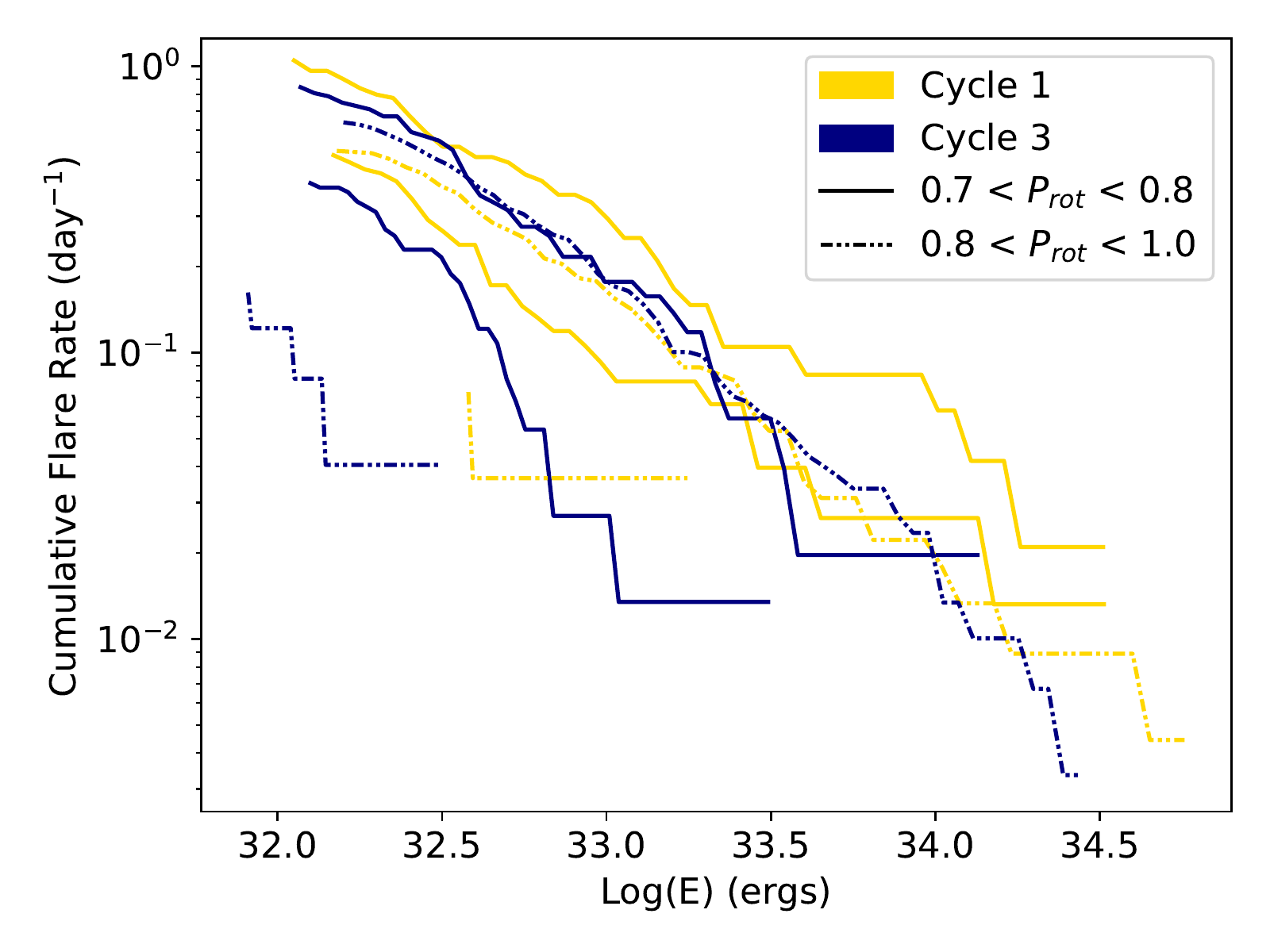} }\\
  \caption{The cumulative flare frequency distribution (FFD) for all 605 flares observed on the ten low mass UFR stars. This has been split into Cycle 1 (red) and Cycle 3 (blue) to allow for a direct comparison between the cycles. The top plot is the first three period bins with $P_{\rm rot}$ < 0.6~d and the bottom the remaining two period bins with 0.7 < $P_{\rm rot}$ < 1.0~d. The flares have been sorted from smallest to largest by their energy per day. }
    \label{flare_freq}
    \end{center}
\end{figure}

We now consider whether the flare rate as a function of energy changed between the two Cycles. In Figure \ref{flare_freq}  we show the Flare Frequency Distribution (FFD) which shows the occurrence rate of flares from the stars as a function of energy. This is the most appropriate method to compare levels of flare activity between observations of different lengths \citep[see][and references within]{davenport202010years}. It is common to model a stars FFD using a power law distribution \citep[for examples see][]{hawley2014kepler, davenport202010years}. Stars with significant differences in the power law slope would indicate changes in the relative energy of flares. However, for stars with relatively small numbers of flares, it is more appropriate to use other statistical tests to determine the difference in the FFD between Cycle 1 and 3. 

We have used the Anderson-Darling and Kolmogorov-Smirnov two sample tests. We find that there are no significant changes in the FFD for any of the stars in our sample between the two cycles (we considered the five stars which show at least 10 flares in each cycle). In each case the random chance probability of the observed difference in energy distributions is $p>$0.045. Next we pooled all the flares into two groups (each including five stars), originating in either i) stars with period $<$0.5 d and ii) stars with period $>$0.5 d. We find that initially there is a clear difference in the FFD between these two groups, but closer inspection reveals it is due to a single source (TIC 22914229) which shows systematically lower flare energies than the rest. Ignoring this single source, we find that there is no significant difference in the flare energy distribution between the shorter period sources and the longer period sources. Further observations of these targets in subsequent {\tess} Cycles will allow searches for longer term variations in activity levels. 

Although we find no evidence for a significant change in the flare rate between different Cycles, we now examine whether the morphology of light curves of the same star can change over the two Cycles.  In Appendix \ref{425937691_sectors} we show a short section of the light curve of UCAC3~53-724 (TIC 425937691) obtained in Sector 1 and 28 along with the flare with the most prominent maximum. The peak in the power spectra is 0.20 d which we take to be the rotation period. Despite the flare rate showing no significant variation, there is a clear difference in the shape of the light curve between sectors. In Sector 1 it is possible to identify peaks in the light curve every $\sim$0.2 d. In Sector 28 there are no clear peaks at this period, rather dips. UCAC3~53-724 is also included in the study of \cite{zhan2019complex} who found ten late type stars which appear to be rapidly rotating but also show dips in their lightcurve which could be caused by material close to the star, possibly in a dusty ring structure. We conclude that although the distribution of spots/active areas may have changed between cycles it does not significantly change the flare rate. This is somewhat expected, as the amplitude of modulation generated by the spots remains similar in both cycles, implying comparable spot coverage overall.

AL~442 (TIC 141807839), observed near the continuous viewing zone (CVZ), is a second comparison star. There are 21 sectors of data for this star, to date, which approximates to $\sim$465~d of observations over 3~years. We show sections from Cycles 1 and 3 of the 2-min cadence {\tess} lightcurves for this target in Appendix \ref{141807839_cycles}. Again, despite the lack of significant change in the flare rate we find  a change in the shape of the rotational modulation of the lightcurve indicating changes in the distribution/number of spots/active regions. Specifically, in Cycle 1, the rotational modulation is more regular with a repeating sinusoidal shape with the same amplitude every $\sim$0.8~d. In contrast, for Cycle 3, the modulation is very irregular over the $\sim$0.8~d rotation period with varying sinusoidal shapes with different amplitudes. We know sunspots migrate to latitudes closer to the equator over a Solar activity cycle. Therefore, this could explain the change in the rotational modulation, if the existing spots/active regions have moved closer to the equator. As a result, this change in latitude of the spots/active regions will alter the shape of the modulation within the lightcurve but not changing the flare rate of the star. 

We conclude that although the flare rate does not change over the several year time scale, the shapes of the rotational modulation for the two targets we have discussed are different. The total spot coverage does not change (i.e. the amplitude of the modulation remains approximately the same) but the spot distribution can change between the {\tess} Cycles.

\section{Discussion \& Conclusion}
\label{discussion}

Studies of low mass stars made using {\tess} data have revealed a sample of stars which have light curves which show a periodic signal at periods $<$0.2 d which show no or few flares. If these periods represent the rotation period of the star then this is surprising since we expect rapidly rotating stars to be very active. In this paper we sought out to determine if a sample of ten low mass UFRs (spectral types between M2--M6V) with a periodic modulation $<$0.2 d showed any evidence for a magnetic field. To do this we utilised VLT/FORS2 spectro-polarimetric observations and determined that five out of the ten stars have a detectable magnetic field of $\sim$1-2 kG. Comparing {\tess} data taken during Cycle 1 and 3 showed that none of the stars showed a significant change in the number of flares between Cycles, but two stars showed very different light curve morphologies.

Two of our targets, UCAC3 53-724 (TIC 425937691) and UPM~J0113-5939 (TIC 206544316), were present in the study of \citet{zhan2019complex} who investigate the complex and highly structured nature of the rotational modulation for ten low mass M (M3.5 -- M6.5V) dwarfs. In their study, they compare their sample of M dwarfs to the rapidly rotating and flare active M dwarf V374~Peg which is expected to have a stable global poloidal magnetic field of between 1.5--2~kG. As a result of this, they expect UPM~J0113-5939 also has a strong stable global field comparable to V374~Peg. In our study we detect a magnetic field on UPM~J0113-5939 with a strength of $\sim$0.7~kG (no field was detected on UCAC3 53-724), more than half that of V374~Peg. \citet{zhan2019complex} conclude the highly structured nature of the rotational modulation of UPM~J0113-5939 is a result of a dusty ring surrounding the star. Currently the nature of young M dwarfs with complex, sharp peaked and periodic rotational modulation is unknown with several scenarios to explain the origins including: spots only; accreting dust disks; co-rotating clouds of material; magnetically constrained material; spots, and a spin-orbit-misaligned disk, and spots and pulsations \citep[see][for more details]{stauffer2017orbiting, gunther2020complex, koen2021multifilter}.

With only half of our sample having a detectable magnetic field and four of those being the more flare active stars in the sample, it would appear the magnetic field strength may not be the answer to the lack of flaring activity in UFRs: it remains to be seen whether the field configuration is a factor (i.e. by using Doppler imaging to infer this). Interestingly, we do not detect a magnetic field on our second fastest rotating star UCAC3~53-724 which has a rotation period of 0.2~d. Further examination shows a change in the shape of the rotational modulation, which can be seen in Appendix \ref{425937691_sectors}, indicating a change in the number and/or distribution of spots/active regions on the stellar surface. According to the rotation-activity-age relation, young stars rotate faster and, therefore, produce higher levels of flaring activity. 

\citet{medina2020flare} conducted an analysis into the flare rates, rotation periods, and spectroscopic activity indicators of 125 single low mass stars within 15~pc, observed during the first year of the {\tess} mission. They find a clear saturated relationship between the flare rate and Rossby number (rotation period of star divided by its convective turnover time), where flares per day decreases rapidly with increasing Rossby number and rotation period, consistent with what was observed in this small group of UFRs. However, they do not place age estimates on their sample and state  the relationship between age and activity may be more complicated than a steady decrease over time due to the complexities surrounding how stars shed angular momentum. For our sample, we were able to use multiple studies (see Table \ref{stellar_properties}) to place age estimates on seven of our targets since they are members of moving groups or associations. These ages range between 13 -- 150~Myr which would place them in an age group which is young and active \citep[e.g.][]{Kiman2021}.

The TESS lightcurves clearly show that the rotational modulation shape evolves with time. Overall, this can be understood with the presence of evolving spots on the stellar surface. However, highly structured changes in modulation shape can also be attributed to a spotted star surrounded by a dusty ring  \citep[see][]{zhan2019complex}. We have detected a measurable magnetic field in half of the stars studied here. However, the detection limit is such, that we cannot rule out spots on the stars with magnetic field strengths just below that limit. Therefore, it is viable that all the stars observed here do show magnetic activity. This is also expected, since if the photometric periods reflect the rotation periods of the stars, they should have an operational magnetic dynamo within their convective layer, even in the case of fully convective, very late type M dwarfs \citep{2006ApJ...638..336D}.

However, if all of these sources show magnetic activity, then why do some of them, especially the shortest period ones \citep[e.g.][]{ramsay2020ufr}, do not show any flares? We speculate that this might be related to the well known phenomenon of supersaturation that has been observed in the X-ray activity of fast rotating M dwarfs \citep{2011MNRAS.411.2099J}. According to \citet{2011MNRAS.411.2099J}, when the stellar rotation increases enough, the Keplerian co-rotation radius moves inwards and can reside within the stellar corona. This would lead to the centrifugal forces ripping up the upper corona and opening the magnetic loops within that region. This appears to be the case for M dwarfs with P$<$0.2~d \citep{2011MNRAS.411.2099J} with the centrifugal force being a factor of four stronger at P=0.15~d compared to P=0.3~d. This would naturally imply at least a reduced rate of magnetic field reconnection, which in turn, would ultimately lead to much reduced or maybe even totally inhibited flaring. It could then be possible that we would witness magnetic field related spot activity on the stellar photosphere, whilst not detecting any flares. 

In a companion paper \citep{ramsay2021ufr} we present a study made using data obtained from the Nordic Optical Telescope where we searched for radial velocity variations in 29 stars which lie close to the lower main sequence in the Gaia HRD. There is only one star which shows clear evidence for radial velocity variations, with another two stars showing high Gaia RUWE values indicating they are binaries. An additional four stars show folded {\tess} lightcurves which indicate they are either short period pulsating stars or eclipsing binaries. The conclusion is that the lack of flares in UFR is unlikely to be explained by the stars being members of binaries or that the modulation is due to blending with other variable stars.

\section*{Acknowledgements}

This work is based on observations made with ESO Telescopes at the La Silla Paranal Observatory under the programme ID 106.210N.001. We also include data collected by the {\tess} mission, where funding for the {\tess} mission is provided by the NASA Explorer Program. In addition, we present results from the European Space Agency (ESA) space mission {\sl Gaia}. The {\sl Gaia} data was processed by the {\sl Gaia} Data Processing and Analysis Consortium (DPAC). Funding for the DPAC is provided by national institutions, in particular the institutions participating in the {\sl Gaia} MultiLateral Agreement (MLA). The Gaia mission website is \url{https://www.cosmos.esa.int/gaia} and data can be accessed from \url{https://archives.esac.esa.int/gaia}. LD acknowledges funding from a UKRI Future Leader Fellowship, grant number MR/S035214/1. Armagh Observatory and Planetarium is core funded by the N. Ireland Department of Communities. JGD would like to thank the Leverhulme Trust for a Emeritus Fellowship. This work made use of \texttt{tpfplotter} by J. Lillo-Box (publicly available in \url{www.github.com/jlillo/tpfplotter}), which also made use of the python packages \texttt{astropy}, \texttt{lightkurve}, \texttt{matplotlib} and \texttt{numpy}. Finally, we would like to thank Axel Brandenburg for some valuable comments regarding magnetic dynamo generation, and an anonymous referee for a helpful and constructive report.

\section*{Data Availability}
The {\tess} data are available from the NASA MAST portal and the ESO VLT/FORS2 data are public from the ESO data archive.



\bibliographystyle{mnras}
\bibliography{magfields_ufr} 

\begin{thebibliography}{}
\makeatletter
\relax
\def\mn@urlcharsother{\let\do\@makeother \do\$\do\&\do\#\do\^\do\_\do\%\do\~}
\def\mn@doi{\begingroup\mn@urlcharsother \@ifnextchar [ {\mn@doi@}
  {\mn@doi@[]}}
\def\mn@doi@[#1]#2{\def\@tempa{#1}\ifx\@tempa\@empty \href
  {http://dx.doi.org/#2} {doi:#2}\else \href {http://dx.doi.org/#2} {#1}\fi
  \endgroup}
\def\mn@eprint#1#2{\mn@eprint@#1:#2::\@nil}
\def\mn@eprint@arXiv#1{\href {http://arxiv.org/abs/#1} {{\tt arXiv:#1}}}
\def\mn@eprint@dblp#1{\href {http://dblp.uni-trier.de/rec/bibtex/#1.xml}
  {dblp:#1}}
\def\mn@eprint@#1:#2:#3:#4\@nil{\def\@tempa {#1}\def\@tempb {#2}\def\@tempc
  {#3}\ifx \@tempc \@empty \let \@tempc \@tempb \let \@tempb \@tempa \fi \ifx
  \@tempb \@empty \def\@tempb {arXiv}\fi \@ifundefined
  {mn@eprint@\@tempb}{\@tempb:\@tempc}{\expandafter \expandafter \csname
  mn@eprint@\@tempb\endcsname \expandafter{\@tempc}}}

\bibitem[\protect\citeauthoryear{{Afrin Badhan}, {Wolf}, {Kopparapu}, {Arney},
  {Kempton}, {Deming}  \& {Domagal-Goldman}}{{Afrin Badhan}
  et~al.}{2019}]{Badhan2019}
{Afrin Badhan} M.,  {Wolf} E.~T.,  {Kopparapu} R.~K.,  {Arney} G.,  {Kempton}
  E. M.~R.,  {Deming} D.,   {Domagal-Goldman} S.~D.,  2019, \mn@doi [\apj]
  {10.3847/1538-4357/ab32e8}, \href
  {https://ui.adsabs.harvard.edu/abs/2019ApJ...887...34A} {887, 34}

\bibitem[\protect\citeauthoryear{{Aller}, {Lillo-Box}, {Jones}, {Miranda}  \&
  {Barcel{\'o} Forteza}}{{Aller} et~al.}{2020}]{Aller2020}
{Aller} A.,  {Lillo-Box} J.,  {Jones} D.,  {Miranda} L.~F.,   {Barcel{\'o}
  Forteza} S.,  2020, \mn@doi [\aap] {10.1051/0004-6361/201937118}, \href
  {https://ui.adsabs.harvard.edu/abs/2020A&A...635A.128A} {635, A128}

\bibitem[\protect\citeauthoryear{Appenzeller \& Rupprecht}{Appenzeller \&
  Rupprecht}{1992}]{appenzeller1992fors}
Appenzeller I.,  Rupprecht G.,  1992, The Messenger, 67, 18

\bibitem[\protect\citeauthoryear{Appenzeller et~al.,}{Appenzeller
  et~al.}{1998}]{appenzeller1998fors}
Appenzeller I.,  et~al., 1998, The Messenger, 94

\bibitem[\protect\citeauthoryear{{Asensio Ramos}, {Mart{\'\i}nez Gonz{\'a}lez},
  {Manso Sainz}, {Corradi}  \& {Leone}}{{Asensio Ramos}
  et~al.}{2014}]{AsensioRamos2014}
{Asensio Ramos} A.,  {Mart{\'\i}nez Gonz{\'a}lez} M.~J.,  {Manso Sainz} R.,
  {Corradi} R.~L.~M.,   {Leone} F.,  2014, \mn@doi [\apj]
  {10.1088/0004-637X/787/2/111}, \href
  {https://ui.adsabs.harvard.edu/abs/2014ApJ...787..111A} {787, 111}

\bibitem[\protect\citeauthoryear{{Bagnulo} \& {Landstreet}}{{Bagnulo} \&
  {Landstreet}}{2018}]{BagLan18}
{Bagnulo} S.,  {Landstreet} J.~D.,  2018, \mn@doi [\aap]
  {10.1051/0004-6361/201833235}, \href
  {http://adsabs.harvard.edu/abs/2018A%26A...618A.113B} {618, A113}

\bibitem[\protect\citeauthoryear{{Bagnulo}, {Landolfi}, {Landstreet}, {Landi
  Degl'Innocenti}, {Fossati}  \& {Sterzik}}{{Bagnulo} et~al.}{2009}]{Bagetal09}
{Bagnulo} S.,  {Landolfi} M.,  {Landstreet} J.~D.,  {Landi Degl'Innocenti} E.,
  {Fossati} L.,   {Sterzik} M.,  2009, \mn@doi [\pasp] {10.1086/605654}, \href
  {http://adsabs.harvard.edu/abs/2009PASP..121..993B} {121, 993}

\bibitem[\protect\citeauthoryear{{Bagnulo} et~al.,}{{Bagnulo}
  et~al.}{2020}]{Bagnulo2020}
{Bagnulo} S.,  et~al., 2020, \mn@doi [\aap] {10.1051/0004-6361/201937098},
  \href {https://ui.adsabs.harvard.edu/abs/2020A&A...635A.163B} {635, A163}

\bibitem[\protect\citeauthoryear{Booth, del Burgo  \& Hambaryan}{Booth
  et~al.}{2021}]{booth2021age}
Booth M.,  del Burgo C.,   Hambaryan V.~V.,  2021, \mnras, 500, 5552

\bibitem[\protect\citeauthoryear{Davenport}{Davenport}{2016}]{davenport2016kepler}
Davenport J.~R.,  2016, \apj, 829, 23

\bibitem[\protect\citeauthoryear{Davenport et~al.,}{Davenport
  et~al.}{2014}]{davenport2014kepler}
Davenport J.~R.,  et~al., 2014, \apj, 797, 122

\bibitem[\protect\citeauthoryear{Davenport, Covey, Clarke, Boeck, Cornet  \&
  Hawley}{Davenport et~al.}{2019}]{davenport2019evolution}
Davenport J.~R.,  Covey K.~R.,  Clarke R.~W.,  Boeck A.~C.,  Cornet J.,
  Hawley S.~L.,  2019, \apj, 871, 241

\bibitem[\protect\citeauthoryear{Davenport, Mendoza  \& Hawley}{Davenport
  et~al.}{2020}]{davenport202010years}
Davenport J.~R.,  Mendoza G.~T.,   Hawley S.~L.,  2020, \aj, 160, 36

\bibitem[\protect\citeauthoryear{Dmitrienko \& Savanov}{Dmitrienko \&
  Savanov}{2017}]{dmitrienko2017spots}
Dmitrienko E.,  Savanov I.,  2017, \arep, 61, 122

\bibitem[\protect\citeauthoryear{{Dobler}, {Stix}  \& {Brandenburg}}{{Dobler}
  et~al.}{2006}]{2006ApJ...638..336D}
{Dobler} W.,  {Stix} M.,   {Brandenburg} A.,  2006, \mn@doi [\apj]
  {10.1086/498634}, \href
  {https://ui.adsabs.harvard.edu/abs/2006ApJ...638..336D} {638, 336}

\bibitem[\protect\citeauthoryear{Doyle, Ramsay, Doyle, Wu  \& Scullion}{Doyle
  et~al.}{2018}]{doyle2018investigating}
Doyle L.,  Ramsay G.,  Doyle J.~G.,  Wu K.,   Scullion E.,  2018, \mnras, 480,
  2153

\bibitem[\protect\citeauthoryear{Doyle, Ramsay, Doyle  \& Wu}{Doyle
  et~al.}{2019}]{doyle2019}
Doyle L.,  Ramsay G.,  Doyle J.~G.,   Wu K.,  2019, \mnras, 489, 437

\bibitem[\protect\citeauthoryear{Feinstein, Montet, Ansdell, Nord, Bean,
  G{\"u}nther, Gully-Santiago  \& Schlieder}{Feinstein
  et~al.}{2020}]{feinstein2020flare}
Feinstein A.~D.,  Montet B.~T.,  Ansdell M.,  Nord B.,  Bean J.~L.,
  G{\"u}nther M.~N.,  Gully-Santiago M.~A.,   Schlieder J.~E.,  2020, \apj,
  160, 219

\bibitem[\protect\citeauthoryear{Gagn{\'e}, Lafreni{\`e}re, Doyon, Malo  \&
  Artigau}{Gagn{\'e} et~al.}{2014}]{gagne2014banyan}
Gagn{\'e} J.,  Lafreni{\`e}re D.,  Doyon R.,  Malo L.,   Artigau {\'E}.,  2014,
  \apj, 798, 73

\bibitem[\protect\citeauthoryear{Gaia~Collaboration}{Gaia~Collaboration}{2016}]{gaia16}
Gaia~Collaboration Brown A. G. A. e.~a.,  2016, \aap, 595, A2

\bibitem[\protect\citeauthoryear{Gaia~Collaboration}{Gaia~Collaboration}{2018}]{gaia18}
Gaia~Collaboration Brown A. G. A. e.~a.,  2018, \aap, 616, A1

\bibitem[\protect\citeauthoryear{Gaia~Collaboration}{Gaia~Collaboration}{2021}]{gaiaedr3brown}
Gaia~Collaboration Brown A. G. A. e.~a.,  2021, \aap, 649, A1

\bibitem[\protect\citeauthoryear{Gregory}{Gregory}{2005}]{gregory2005bayesian}
Gregory P.,  2005, Bayesian Logical Data Analysis for the Physical Sciences: A
  Comparative Approach with Mathematica{\textregistered} Support.
Cambridge University Press

\bibitem[\protect\citeauthoryear{G{\"u}nther et~al.,}{G{\"u}nther
  et~al.}{2020a}]{gunther2020complex}
G{\"u}nther M.~N.,  et~al., 2020a, arXiv preprint arXiv:2008.11681

\bibitem[\protect\citeauthoryear{G{\"u}nther et~al.,}{G{\"u}nther
  et~al.}{2020b}]{gunther2020stellar}
G{\"u}nther M.~N.,  et~al., 2020b, \apj, 159, 60

\bibitem[\protect\citeauthoryear{Hartmann \& Noyes}{Hartmann \&
  Noyes}{1987}]{hartmann1987rotation}
Hartmann L.~W.,  Noyes R.~W.,  1987, \aapr, 25, 271

\bibitem[\protect\citeauthoryear{Hawley, Davenport, Kowalski, Wisniewski, Hebb,
  Deitrick  \& Hilton}{Hawley et~al.}{2014}]{hawley2014kepler}
Hawley S.~L.,  Davenport J.~R.,  Kowalski A.~F.,  Wisniewski J.~P.,  Hebb L.,
  Deitrick R.,   Hilton E.~J.,  2014, \apj, 797, 121

\bibitem[\protect\citeauthoryear{Howard et~al.,}{Howard
  et~al.}{2020}]{howard2020evryflare}
Howard W.~S.,  et~al., 2020, \apj, 895, 140

\bibitem[\protect\citeauthoryear{Ilin, Schmidt, Poppenh{\"a}ger, Davenport,
  Kristiansen  \& Omohundro}{Ilin et~al.}{2021}]{ilin2021flares}
Ilin E.,  Schmidt S.~J.,  Poppenh{\"a}ger K.,  Davenport J.~R.,  Kristiansen
  M.~H.,   Omohundro M.,  2021, \aap, 645, A42

\bibitem[\protect\citeauthoryear{Janson, Durkan, Hippler, Dai, Brandner,
  Schlieder, Bonnefoy  \& Henning}{Janson et~al.}{2017}]{janson2017binaries}
Janson M.,  Durkan S.,  Hippler S.,  Dai X.,  Brandner W.,  Schlieder J.,
  Bonnefoy M.,   Henning T.,  2017, \aap, 599, A70

\bibitem[\protect\citeauthoryear{{Jeffries}, {Jackson}, {Briggs}, {Evans}  \&
  {Pye}}{{Jeffries} et~al.}{2011}]{2011MNRAS.411.2099J}
{Jeffries} R.~D.,  {Jackson} R.~J.,  {Briggs} K.~R.,  {Evans} P.~A.,   {Pye}
  J.~P.,  2011, \mn@doi [\mnras] {10.1111/j.1365-2966.2010.17848.x}, \href
  {https://ui.adsabs.harvard.edu/abs/2011MNRAS.411.2099J} {411, 2099}

\bibitem[\protect\citeauthoryear{{Kiman} et~al.,}{{Kiman}
  et~al.}{2021}]{Kiman2021}
{Kiman} R.,  et~al., 2021, \mn@doi [\aj] {10.3847/1538-3881/abf561}, \href
  {https://ui.adsabs.harvard.edu/abs/2021AJ....161..277K} {161, 277}

\bibitem[\protect\citeauthoryear{Kochukhov \& Lavail}{Kochukhov \&
  Lavail}{2017}]{kochukhov2017global}
Kochukhov O.,  Lavail A.,  2017, \apjl, 835, L4

\bibitem[\protect\citeauthoryear{Koen}{Koen}{2021}]{koen2021multifilter}
Koen C.,  2021, \aj, 162, 2

\bibitem[\protect\citeauthoryear{{Kolenberg} \& {Bagnulo}}{{Kolenberg} \&
  {Bagnulo}}{2009}]{KolenbergBagnulo2009}
{Kolenberg} K.,  {Bagnulo} S.,  2009, \mn@doi [\aap]
  {10.1051/0004-6361/200811591}, \href
  {https://ui.adsabs.harvard.edu/abs/2009A&A...498..543K} {498, 543}

\bibitem[\protect\citeauthoryear{Loyd et~al.,}{Loyd
  et~al.}{2021}]{loyd2021hazmat}
Loyd R.~P.,  et~al., 2021, \apj, 907, 91

\bibitem[\protect\citeauthoryear{Maggio, Sciortino, Vaiana, Majer, Bookbinder,
  Golub, Harnden~Jr  \& Rosner}{Maggio et~al.}{1987}]{maggio1987einstein}
Maggio A.,  Sciortino S.,  Vaiana G.,  Majer P.,  Bookbinder J.,  Golub L.,
  Harnden~Jr F.,   Rosner R.,  1987, \apj, 315, 687

\bibitem[\protect\citeauthoryear{Martins et~al.,}{Martins
  et~al.}{2020}]{martins2020search}
Martins B.~C.,  et~al., 2020, \apjs, 250, 20

\bibitem[\protect\citeauthoryear{McQuillan, Aigrain  \& Mazeh}{McQuillan
  et~al.}{2013}]{mcquillan2013measuring}
McQuillan A.,  Aigrain S.,   Mazeh T.,  2013, \mnras, 432, 1203

\bibitem[\protect\citeauthoryear{Medina, Winters, Irwin  \& Charbonneau}{Medina
  et~al.}{2020}]{medina2020flare}
Medina A.~A.,  Winters J.~G.,  Irwin J.~M.,   Charbonneau D.,  2020, \aj, 905,
  107

\bibitem[\protect\citeauthoryear{Metcalfe et~al.,}{Metcalfe
  et~al.}{2021}]{metcalfe2021magnetic}
Metcalfe T.~S.,  et~al., 2021, The Astrophysical Journal, 921, 122

\bibitem[\protect\citeauthoryear{Mullan \& Houdebine}{Mullan \&
  Houdebine}{2020}]{mullan2020transition}
Mullan D.,  Houdebine E.,  2020, \apj, 891, 128

\bibitem[\protect\citeauthoryear{Newton, Irwin, Charbonneau, Berta-Thompson,
  Dittmann  \& West}{Newton et~al.}{2016}]{newton2016rotation}
Newton E.~R.,  Irwin J.,  Charbonneau D.,  Berta-Thompson Z.~K.,  Dittmann
  J.~A.,   West A.~A.,  2016, \apj, 821, 93

\bibitem[\protect\citeauthoryear{{Nutzman} \& {Charbonneau}}{{Nutzman} \&
  {Charbonneau}}{2008}]{NutzmanCharbonneau2008}
{Nutzman} P.,  {Charbonneau} D.,  2008, \mn@doi [\pasp] {10.1086/533420}, \href
  {https://ui.adsabs.harvard.edu/abs/2008PASP..120..317N} {120, 317}

\bibitem[\protect\citeauthoryear{{Pecaut} \& {Mamajek}}{{Pecaut} \&
  {Mamajek}}{2013}]{PecautMamajek2013}
{Pecaut} M.~J.,  {Mamajek} E.~E.,  2013, \mn@doi [\apjs]
  {10.1088/0067-0049/208/1/9}, \href
  {https://ui.adsabs.harvard.edu/abs/2013ApJS..208....9P} {208, 9}

\bibitem[\protect\citeauthoryear{{Petit} \& {Wade}}{{Petit} \&
  {Wade}}{2012}]{PetitWade2012}
{Petit} V.,  {Wade} G.~A.,  2012, \mn@doi [\mnras]
  {10.1111/j.1365-2966.2011.20091.x}, \href
  {https://ui.adsabs.harvard.edu/abs/2012MNRAS.420..773P} {420, 773}

\bibitem[\protect\citeauthoryear{{Press}, {Teukolsky}, {Vetterling}  \&
  {Flannery}}{{Press} et~al.}{1992}]{NRF1992}
{Press} W.~H.,  {Teukolsky} S.~A.,  {Vetterling} W.~T.,   {Flannery} B.~P.,
  1992, {Numerical recipes in FORTRAN. The art of scientific computing}

\bibitem[\protect\citeauthoryear{Ramsay, Doyle  \& Doyle}{Ramsay
  et~al.}{2020}]{ramsay2020ufr}
Ramsay G.,  Doyle J.~G.,   Doyle L.,  2020, \mnras, 497, 2320

\bibitem[\protect\citeauthoryear{Ramsay, Hakala, Doyle  \& Doyle}{Ramsay
  et~al.}{2021}]{ramsay2021ufr}
Ramsay G.,  Hakala P.,  Doyle J.~G.,   Doyle L.,  2021, \mnras, submitted.

\bibitem[\protect\citeauthoryear{{Ricker} et~al.,}{{Ricker}
  et~al.}{2015}]{Ricker2015}
{Ricker} G.~R.,  et~al., 2015, \mn@doi [Journal of Astronomical Telescopes,
  Instruments, and Systems] {10.1117/1.JATIS.1.1.014003}, \href
  {http://adsabs.harvard.edu/abs/2015JATIS...1a4003R} {1, 014003}

\bibitem[\protect\citeauthoryear{{Schwarzenberg-Czerny}}{{Schwarzenberg-Czerny}}{1996}]{Schwarzenberg-Czerny1996}
{Schwarzenberg-Czerny} A.,  1996, \mn@doi [\apjl] {10.1086/309985}, \href
  {https://ui.adsabs.harvard.edu/abs/1996ApJ...460L.107S} {460, L107}

\bibitem[\protect\citeauthoryear{Seli, Vida, Mo{\'o}r, P{\'a}l  \&
  Ol{\'a}h}{Seli et~al.}{2021}]{seli2021activity}
Seli B.,  Vida K.,  Mo{\'o}r A.,  P{\'a}l A.,   Ol{\'a}h K.,  2021, \aap, 650,
  A138

\bibitem[\protect\citeauthoryear{{Skumanich}}{{Skumanich}}{1972}]{Skumanich1972}
{Skumanich} A.,  1972, \mn@doi [\apj] {10.1086/151310}, \href
  {https://ui.adsabs.harvard.edu/abs/1972ApJ...171..565S} {171, 565}

\bibitem[\protect\citeauthoryear{{Somers} \& {Stassun}}{{Somers} \&
  {Stassun}}{2017}]{SomersStassun2017}
{Somers} G.,  {Stassun} K.~G.,  2017, \mn@doi [\aj]
  {10.3847/1538-3881/153/3/101}, \href
  {https://ui.adsabs.harvard.edu/abs/2017AJ....153..101S} {153, 101}

\bibitem[\protect\citeauthoryear{Stassun \& Torres}{Stassun \&
  Torres}{2021}]{stassun2021parallax}
Stassun K.~G.,  Torres G.,  2021, \apjl, 907, L33

\bibitem[\protect\citeauthoryear{{Stassun}, {Kratter}, {Scholz}  \&
  {Dupuy}}{{Stassun} et~al.}{2012}]{Stassun2012}
{Stassun} K.~G.,  {Kratter} K.~M.,  {Scholz} A.,   {Dupuy} T.~J.,  2012,
  \mn@doi [\apj] {10.1088/0004-637X/756/1/47}, \href
  {https://ui.adsabs.harvard.edu/abs/2012ApJ...756...47S} {756, 47}

\bibitem[\protect\citeauthoryear{Stassun et~al.,}{Stassun
  et~al.}{2018}]{stassun2018tess}
Stassun K.~G.,  et~al., 2018, \apj, 156, 102

\bibitem[\protect\citeauthoryear{Stauffer et~al.,}{Stauffer
  et~al.}{2017}]{stauffer2017orbiting}
Stauffer J.,  et~al., 2017, \aj, 153, 152

\bibitem[\protect\citeauthoryear{{Sullivan} et~al.,}{{Sullivan}
  et~al.}{2015}]{Sullivan2015}
{Sullivan} P.~W.,  et~al., 2015, \mn@doi [\apj] {10.1088/0004-637X/809/1/77},
  \href {https://ui.adsabs.harvard.edu/abs/2015ApJ...809...77S} {809, 77}

\bibitem[\protect\citeauthoryear{Vida, K{\H{o}}v{\'a}ri, P{\'a}l, Ol{\'a}h  \&
  Kriskovics}{Vida et~al.}{2017}]{vida2017frequent}
Vida K.,  K{\H{o}}v{\'a}ri Z.,  P{\'a}l A.,  Ol{\'a}h K.,   Kriskovics L.,
  2017, \apj, 841, 124

\bibitem[\protect\citeauthoryear{Vida, Ol{\'a}h, K{\H{o}}v{\'a}ri, van
  Driel-Gesztelyi, Mo{\'o}r  \& P{\'a}l}{Vida et~al.}{2019}]{vida2019flaring}
Vida K.,  Ol{\'a}h K.,  K{\H{o}}v{\'a}ri Z.,  van Driel-Gesztelyi L.,  Mo{\'o}r
  A.,   P{\'a}l A.,  2019, \apj, 884, 160

\bibitem[\protect\citeauthoryear{Wright, Drake, Mamajek  \& Henry}{Wright
  et~al.}{2011}]{wright2011stellar}
Wright N.~J.,  Drake J.~J.,  Mamajek E.~E.,   Henry G.~W.,  2011, \apj, 743, 48

\bibitem[\protect\citeauthoryear{Yang et~al.,}{Yang
  et~al.}{2017}]{yang2017flaring}
Yang H.,  et~al., 2017, \apj, 849, 36

\bibitem[\protect\citeauthoryear{{Zechmeister} \& {K{\"u}rster}}{{Zechmeister}
  \& {K{\"u}rster}}{2009}]{Zechmeister2009}
{Zechmeister} M.,  {K{\"u}rster} M.,  2009, \mn@doi [\aap]
  {10.1051/0004-6361:200811296}, \href
  {https://ui.adsabs.harvard.edu/abs/2009A&A...496..577Z} {496, 577}

\bibitem[\protect\citeauthoryear{Zhan et~al.,}{Zhan
  et~al.}{2019}]{zhan2019complex}
Zhan Z.,  et~al., 2019, \apj, 876, 127

\makeatother
\end{thebibliography}

\appendix

\section{{\tess} lightcurves of UCAC3 53-724 and AL~442}

\begin{figure*}
  \begin{center}
  \includegraphics[width=0.97\textwidth]{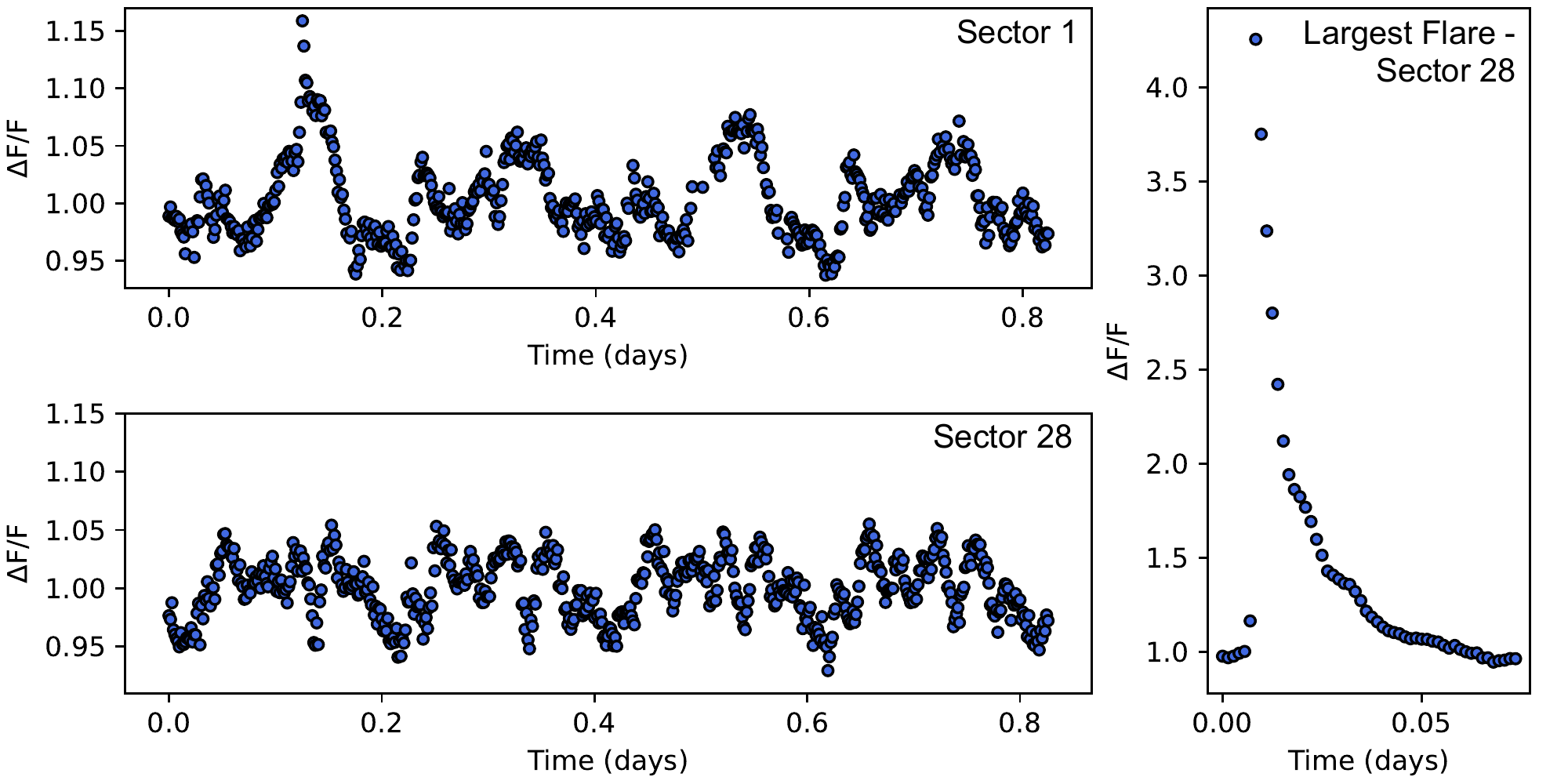}    
  \caption{A selection of portions from {\tess} lightcurves from the M5.5 dwarf UCAC3 53-724 (TIC 425937691). The upper panel shows a section from Sector 1 and the lower panel a section from Sector 28, taken up to 3 years apart. The right hand panel shows the largest flare from this M dwarf with an energy of 1.7 $\times$ 10$^{34}$~erg.}
    \label{425937691_sectors}
    \end{center}
\end{figure*}

\begin{figure*}
  \begin{center}
  \includegraphics[width=0.97\textwidth]{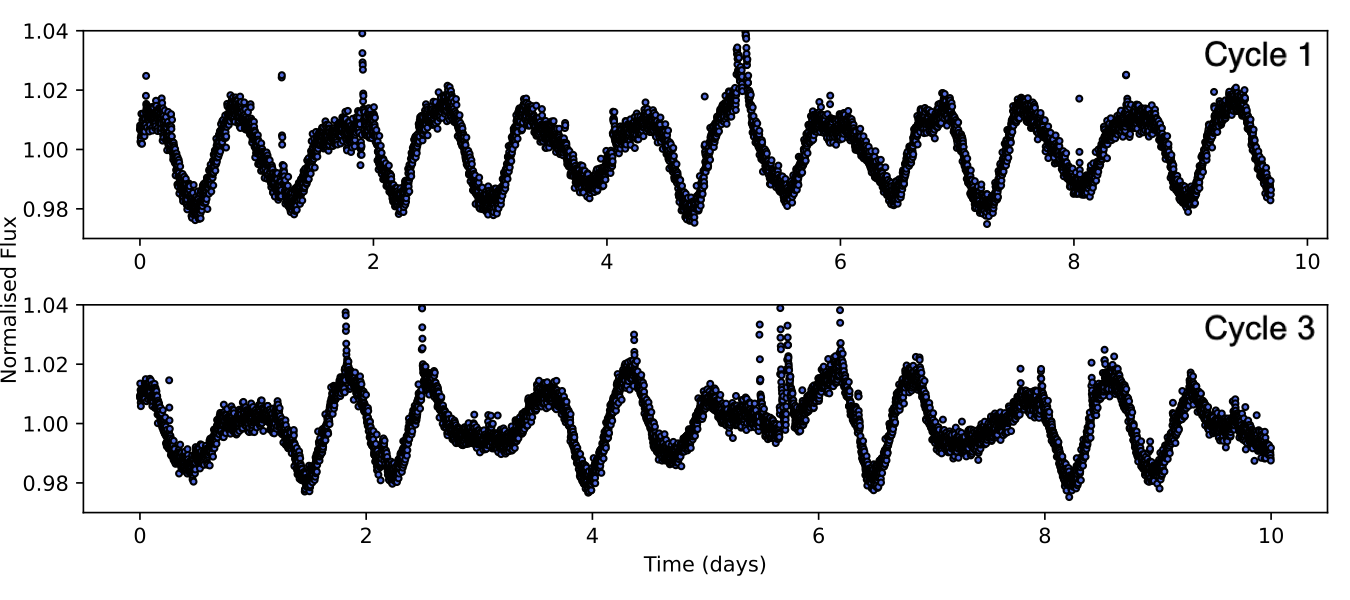}    
  \caption{Two sections of {\tess} lightcurves from Cycle 1 (Sector 11: top panel) and Cycle 3 (Sector 35: bottom panel) of the M4.5 dwarf AL~442 (TIC 141807839). A comparison of the two lightcurves, taken up to 3 years apart, shows a change in the shape of the modulation caused by increased spot activity on the surface of the star. }
    \label{141807839_cycles}
    \end{center}
\end{figure*}

\bsp	
\label{lastpage}
\end{document}